\pdfoutput=1
\documentclass[cits]{JINST}
\newcommand{\unit}[1]{\ensuremath{\mathrm{\,#1}}}
\renewcommand{\u}[1]{\unit{#1}}

\DeclareFontFamily{U}{euc}{}
\DeclareFontShape{U}{euc}{m}{n}{<-6>eurm5<6-8>eurm7<8->eurm10}{}%
\DeclareSymbolFont{AMSc}{U}{euc}{m}{n} 
\DeclareMathSymbol{\umu}{\mathord}{AMSc}{"16}

\title{Performance of Glass Resistive Plate Chambers for a high-granularity semi-digital calorimeter}

\author{
  M. Bedjidian$^a$, K.~Belkadhi$^b$, V.~Boudry$^b$, C.~Combaret$^a$, D.~Decotigny$^b$, E.~Cortina
  Gil$^c$, C.~de la Taille$^d$, R.~Dellanegra$^a$, V.A.~Gapienko$^e$,
  G.~Grenier$^a$, C.~Jauffret$^b$, R.~Kieffer$^a$, M.-C.~Fouz$^f$, R.~Han$^a$,
  I.~Laktineh$^a$\thanks{Corresponding author.}, N.~Lumb$^a$, K.~Manai$^g$,
  S.~Mannai$^{c}$, H.~Mathez$^a$, L.~Mirabito$^a$, J.~Puerta Pelayo$^f$, M.~Ruan$^b$,
  F.~Schirra$^a$, N.~Seguin-Moreau$^d$, W.~Tromeur$^a$, M.~Tytgat$^h$,
  M.~Vander~Donckt$^a$, N.~Zaganidis$^h$\\ 
  \llap{$^a$} Universit\'e de Lyon, Universit\'e Lyon 1, CNRS/IN2P3, 
  IPNL, 4 Rue E.~Fermi, 69622 Villeurbanne Cedex, France\\
  \llap{$^b$}Laboratoire Leprince-Ringuet -- \'Ecole polytechnique, CNRS/IN2P3,
  Palaiseau, F-91128 France\\
  \llap{$^c$}Center for Cosmology Particle Physics and Phenomenology (CP3),
Universit\'{e} Catholique de Louvain, Belgium\\
  \llap{$^d$} Laboratoire de l'Acc\'elerateur Lin\'eaire,
Centre d'Orsay, Universit\'e de Paris-Sud XI,
BP 34, B\^atiment 200, F-91898 Orsay CEDEX, France\\
 \llap{$^e$}Institute of High Energy Physics,
Moscow Region, RU-142284 Protvino, Russia\\
  \llap{$^f$}Centro de Investigaciones Energeticas, Medioambientales y Tecnologicas, Madrid, Spain \\
  \llap{$^g$} Tunis El Manar University, Rommana 1068, Tunis BP 94, Tunisia \\
  \llap{$^h$}Ghent University, Dept. Physics and Astronomy, Proeftuinstraat
  86, B-9000 Gent, Belgium\\
  E-mail: \email{laktineh@in2p3.fr}
}

\abstract{ A new design of highly granular hadronic calorimeter using
  Glass Resistive Plate Chambers (GRPCs) with embedded electronics has been
  proposed for the future International Linear Collider (ILC) experiments. It features a 2-bit threshold semi-digital read-out. Several GRPC prototypes with their electronics have been successfully built and tested in pion beams. The design of these detectors is presented along with the test results on efficiency, pad multiplicity,  stability and reproducibility.  }

\begin{document}


\section{Introduction}

The success of future high-energy experiments intended to investigate physics phenomena in the TeV range will be determined by their ability to precisely measure the energy of jets associated with the production of  bosons such as $W^\pm, Z^0, H^0$.  %
Among the different  proposed methods to obtain high-precision jet energy measurements, one of the most attractive techniques is based on the Particle Flow Algorithm (PFA) approach~\cite{PFA}. %
In this approach, particles are tracked in the different sub-detectors and their energy or momentum is estimated in the most precise sub-detector.
This requires electromagnetic and hadronic calorimeters to have a tracking capacity in addition to their usual functionality.   %
This concept is being implemented within the CALICE Collaboration to design a new generation of sampling hadronic calorimeters with high-granularity in both the transverse and the longitudinal directions.  %
A new generation of electronics has been developed to cope with more than 50 million electronic channels needed for such a high-granularity and yet still compact and hermetic detector.  %
The new electronics are embedded in the detector and daisy-chained.  %
It provides a multi threshold read-out while reducing the power consumption by a factor of 100 using a power pulsing scheme.  %
Different gas detectors are proposed as sensitive media for this new generation of hadronic calorimeter.  %
The Glass Resistive Plate Chamber (GRPC)~\cite{GRPC} is one of the candidates that associates high performance with low production cost.   %
Several developments are ongoing in this direction~\cite{repond}.   %
A prototype of semi-digital detector made of 4 small GRPCs equipped with multi-threshold pulsed electronics was built to validate the concept.  %
It was then exposed to both cosmic rays and pion beams at CERN.  

In this paper we first describe the GRPC developed for this prototype as well as  the new electronic read-out and acquisition systems.   %
We then briefly describe the prototype and the testbeam setup at CERN.  %
Finally, the results obtained with this setup are presented and commented upon.

\section{Glass Resistive Plate Chambers}

The hadronic calorimeters (HCAL) in the future ILC experiments will most probably be located inside the magnetic field. This requires a very compact design with the sensitive medium as thin as possible while providing excellent efficiency and good homogeneity. GRPCs  satisfy all these requirements and are simple and cheap detectors. Unlike Bakelite RPC \cite{santico}, they do not need oil surface treatment to operate and thus avoid the related problems \cite{babar}. 

Several thin GRPCs were built. Their dimensions were chosen to be 33.55$\times$8.35 cm$^2$.  This corresponds to 32$\times$8 read-out channels operated by 4 chained ASICs as will be explained in section~\ref{sec.electronics}.%
\begin{figure}[hh]
\begin{center}
\includegraphics[width=0.8\textwidth, trim=0mm 4.5cm 0mm 6cm,clip=true ]{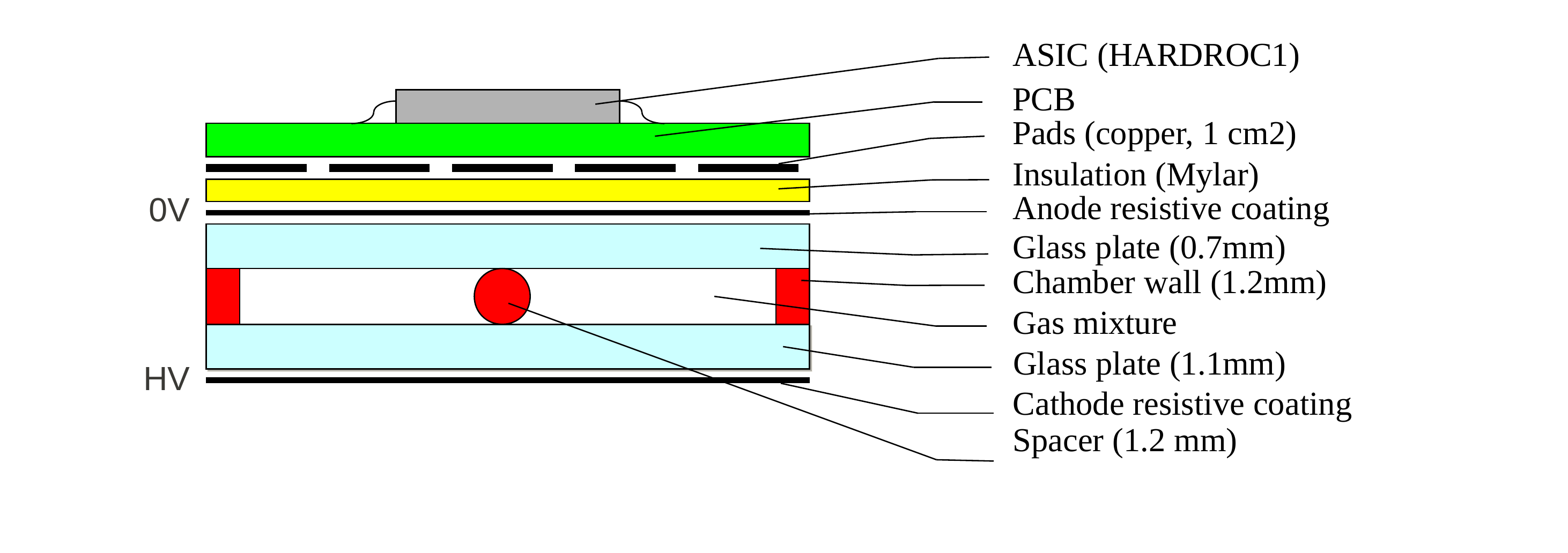}
\end{center}

\caption{Schematic view of a glass RPC\label{fig.rpc}}
\end{figure} 
As shown in Figure \ref{fig.rpc} each GRPC is made of two glass plates.%
The thinner glass (0.7\u{mm} thick) is used to build the anode while the thicker (1.1\u{mm} thick) forms the cathode.   %
The two plates have a resistivity of $10^{12}-10^{13} \Omega$cm. They are kept separated by a 1.2\u{mm}  frame. A fishing line placed longitudinally in the middle of each chamber helps to maintain this distance everywhere inside the chamber.  %
The outer sides of the glass plates are covered by a thin layer of resistive coating and connected to high voltage.  %
A 50\u{\umu m} Mylar$^\copyright{}$  layer separates the anode from the $1\times1\u{cm^2}$ copper pads of the electronic board.  %
The reduced anode thickness is intended to minimise the number of pads with signal (multiplicity).  %
The signal is induced by the charge of the avalanche electrons.  %
The thinner the glass plate  the stronger  is the signal  in the closest pad  and the lower is the relative signal  seen by the neighbouring pads.

A standard graphite-based paint was used for four of these small chambers \cite{protvino}. %
Two other coatings with higher resistivity were used to build additional chambers :  Licron$^\copyright{}$ and Statguard$^\copyright{}$.  %
Table~\ref{tab.coating} summarises the resistive properties of the coatings.

\begin{table}[h]
\begin{center}
\caption{Surface resistivities of glass coatings.\label{tab.coating}}
\begin{tabular}{|l|c|}
\hline
Coating &Resistivity (M$\Omega/\square$)  \\
\hline
Graphite    	&   0.4\\
Statguard  	&   2 \\
Licron       	&  20\\
\hline
\end{tabular}
\end{center}

\end{table}

A gas mixture of tetrafluoroethane (TFE, 93\u{\%}), isobutane (5\u{\%}) and $\mathrm{SF_6}$ (2\u{\%}) is used to operate the GRPC chambers.  %
The TFE commercially known as R134A is used as the ionisation gas with about 8~\u{ionisations / mm} for a MIP while
isobutane and $\mathrm{SF_6}$ are used as gamma and electron quenchers respectively.

The induced charge on the anode in avalanche mode is typically between 0.1 and 10\u{pC} with a rise time of less than 10\u{ns} \cite{protvino}.

\section{Electronics}
\label{sec.electronics}

The future high-granularity HCAL will need more than 50 million read-out channels.  %
It is of primary importance for the future ILC experiments~\cite{ILDproposal} to reduce the power consumption to the lowest level, to be able to cope with such a tremendous amount of channels.  %
A second challenge is to reduce the needed number of external connections and cables.  %
Their effect on the detector geometry can result in many cracks and holes preventing the detector from being hermetic.

\begin{figure}[h]
\begin{center}
\includegraphics[width=0.5\textwidth]{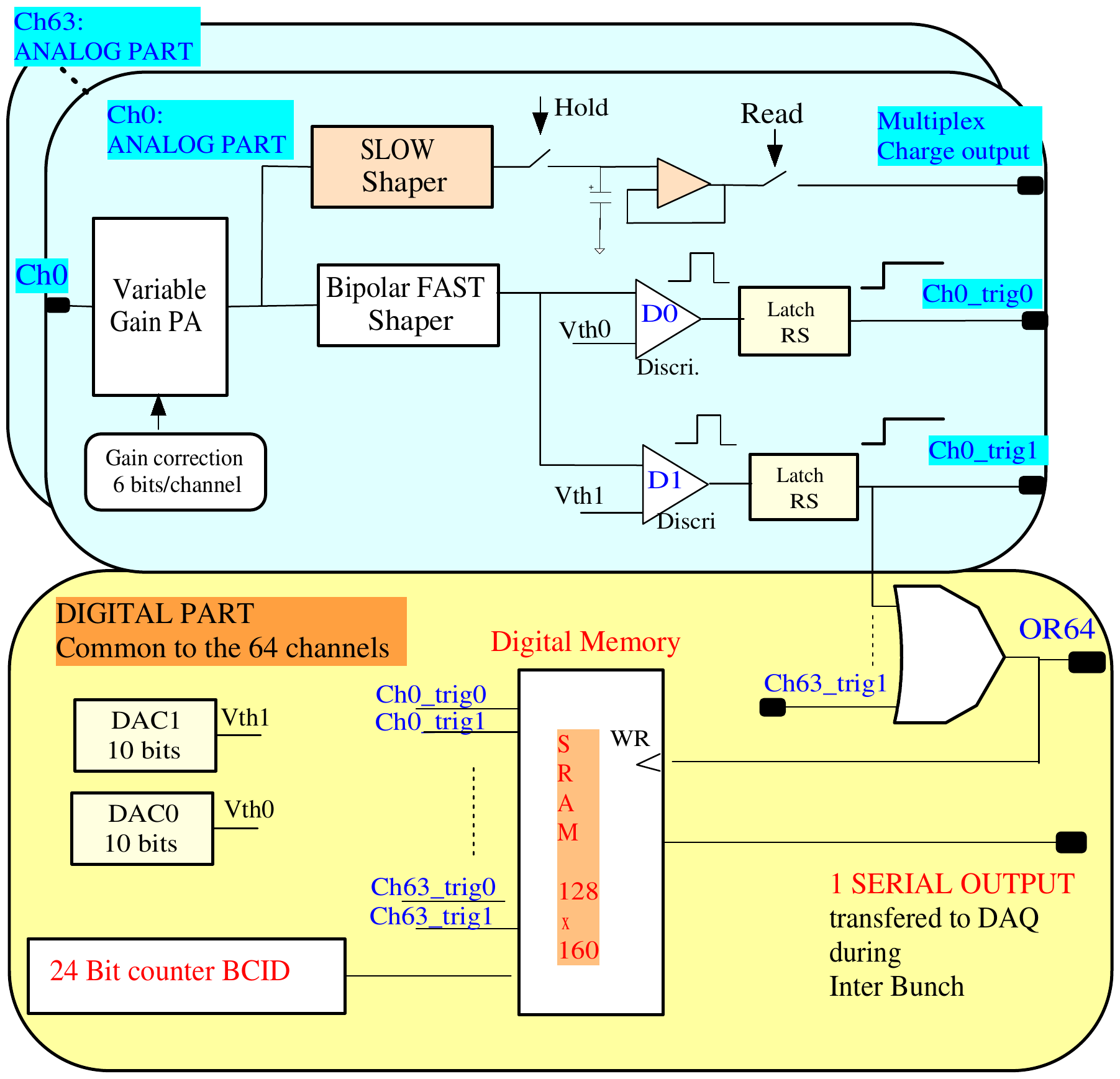}
\caption{Synoptic scheme of the HARDROC chip}
\label{fig.hardroc}
\end{center}
\end{figure}

A specific chip (ASIC) named HARDROC~\cite{hardroc} was developed in AMS (Austrian Micro System) $0.35 \mu m$ Silicium Germanium technology to address these two issues. %
The simplified schematic is displayed in Figure~\ref{fig.hardroc}. %
One ASIC digitises and stores 64 channels.  Each of the 64 channels is made of a fast low-input-impedance current-preamplifier to minimize the crosstalk. The gain of each preamplifier can be varied over 6 bits to correct for the non-uniformity between channels. %
A slow variable shaper (50-150ns) followed by a Track and Hold buffer provides a multiplexed charge measurement up to 10pC very convenient for debug measurements. %
The 2-bit readout is performed using the fast channel which is made of a variable fast shaper (15-50 ns) followed by two low-offset discriminators. %
The comparator's DC reference level (threshold) is set by two 10-bit DACs which are integrated inside the ASIC. Their output value can be configured using integer values in the interval between 0 and 1023 (DAC units).  %
Each of these two thresholds corresponds to an amount of integrated charge on the input.  %
There are 571 Slow Control (SC) parameters which are loaded serially to set the various parameters of the chip. %
Every 200\u{ns} the status of the 64 lowest comparators is evaluated.  %
If one of them is fired, the data are stored in an integrated digital memory; we call this mechanism "auto-triggering".

An important feature of the ASIC is its ability to be power-pulsed.  %
This allows the most power-hungry parts of the chip to be active just before the bunch crossing and off just after.  %
The integrated power consumption is then reduced to less than 10\u{\mu W /channel} in the case of the proposed ILC duty cycle: one millisecond bunch crossing every 200\u{ms}. 
 
The ASIC has been designed to fit in a package small enough to be embedded in the detector.  %
Up to one hundred ASICs of the same detector plane can communicate with each other via a daisy chain using a token ring protocol.  %
Configuration parameters (Slow Control parameters) as well as data collected in the sensor circulate among the ASICs.  %
The readout is sequential and the associated signals (StartReadOut and EndReadOut) go from one ASIC to the next one. %
Hence, the connection with the outside acquisition system is reduced.  %
In case of a dead chip in the daisy chain, it can be bypassed on the printed board using straps or switches to connect the serial Slow Control data line and the Start and End readout signals to the next chip. 
In addition to these features, the ASIC can store up to 128 frames in its internal memory. A frame stored in one ASIC consists of a hit map of the 64 channels  with 2 bits per channel plus a time stamp of 24 bits.  The Gray-coded time stamp is provided by a 5\u{MHz} clock built into the ASICs.  %
This memory is used to record the events which occur during a machine spill (including thousands of bunch crossings) and read them out during the idle time (between spills).  %
Therefore no event will be lost because of online selection.   %
This design relies on the low-occupancy rate (including noise) expected in the HCAL of the future ILC experiments.   %

Another feature of the HARDROC ASIC is the possibility to adjust the gain of each channel separately, by a factor between 0 and 4 with a 6-bit precision.  %
Each channel of the ASIC has a test capacitor of $2\pm0.02\u{pF}$ which can be used to calibrate its response.  %
This is a useful tool to make the response of the different channels as uniform as possible.  %
The cross-talk between two channels of one ASIC was measured by injecting an electric signal equivalent to a MIP (1\u{pC}) through one channel.  %
The signal observed in the other channels was found to be less than 2\u{\%} of the injected charge.

A Printed Circuit Board (PCB) was designed to validate the concept of a semi-digital hadronic calorimeter with high granularity.   %
Hosting 4 ASICs, it provides the connection between adjacent ASICs and the FPGA ensuring the read-out, control and external connectivity through a USB device.  %

The PCB features eight layers, two of them reserved for the digital signal transfer.  %
256 copper pads of $1\times1\u{cm^2}$ were printed on one of the two external PCB faces.   %
The distance between two adjacent pads was chosen to be 500\u{\umu m}.  %
The spread of the avalanches at the anode level is expected to be 1-2\u{mm^2}.  %
This allows to detect even the charged particles which cross the detector between two pads.  %
The routing was optimised to reduce the cross-talk among adjacent pads.   %
A PCB thickness of 800 microns was achieved to minimise the effective detector thickness. 

The pads are connected internally to the ASIC channels through the PCB structure.  %
The cross-talk among adjacent pads was tested before other electronics components were fixed on the PCB:   %
the charges induced on the adjacent pads were measured and found to be less than 0.3\u{\%} of the injected charge which means that the cross-talk due to the PCB is negligible with respect to the ASIC cross-talk mentioned above.

\section{Data acquisition}
\label{sec.acquisition}

An acquisition software developed under the LabView\copyright{} environment uploads the configuration parameters to the different ASICs, manages the ASIC running modes and collects data from these ASICs through the FPGA device. %
Two read-out modes, both using the auto-trigger capacity of the ASIC, were implemented:  %
\begin{itemize}
\item a ``train mode'' (or ILC-like) starts the acquisition on an external
  start-of-spill signal, and stops it either on an end-of-spill or memory-full
  signals.  The events are then read from the the ASIC memory.  This
  mode is foreseen at the ILC.
\item a ``triggered mode'' was conceived for cosmic rays and test beam studies: the acquisition and data taking start on request and are stopped by external triggers. %
  The memory of the ASICs is then read out.  
  If the memory of one of the ASICs is full, a busy signal blocking the external triggers is emitted while the ASICs are reset.  %
  The acquisition resumes after a dead time of two clock counts (400\u{ns}). 
  A 40\u{MHz} external counter provides the time difference between the last recorded frame on the board and the external trigger.  %
This allows for time reconstruction of the occurrence of each frame with respect to the external trigger. 
\end{itemize}
 The read-out lasts between 10 to 75\u{ms} depending on the number of frames in memory.  

\section{Gain correction procedure}
Multi-threshold read-out will be used to distinguish denser regions of the hadronic shower that correspond to the electromagnetic component. Uniform response is therefore important to achieve compensation.

\begin{figure}[h]
 \begin{center}
 \begin{minipage}{0.475\textwidth}
   \begin{centering}
     \includegraphics[width=1\textwidth]{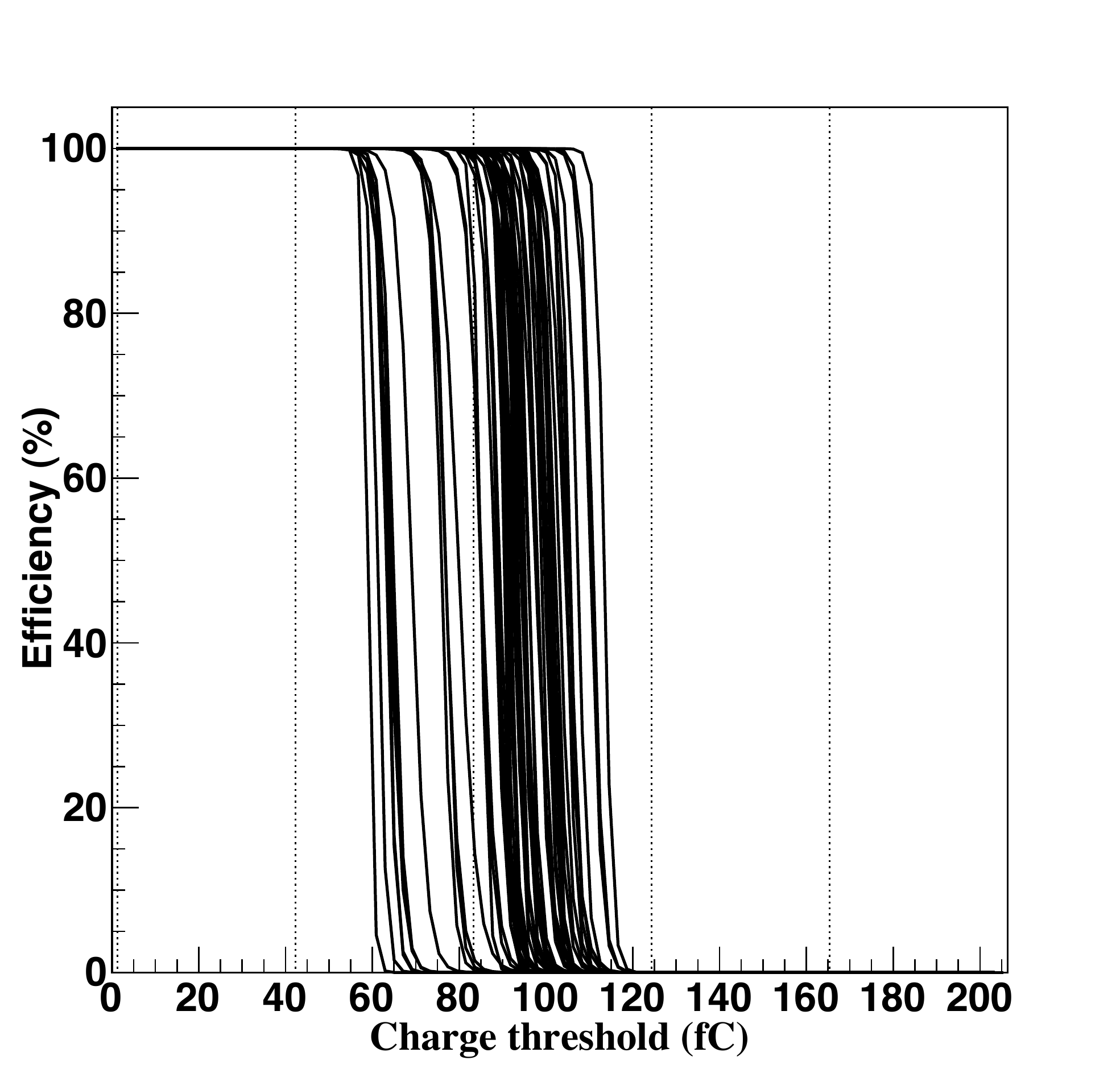}
   \end{centering}
 \end{minipage}
 \begin{minipage}{0.05\textwidth}
 \end{minipage}
 \begin{minipage}{0.475\textwidth}
   \begin{centering}
    \includegraphics[width=1\textwidth]{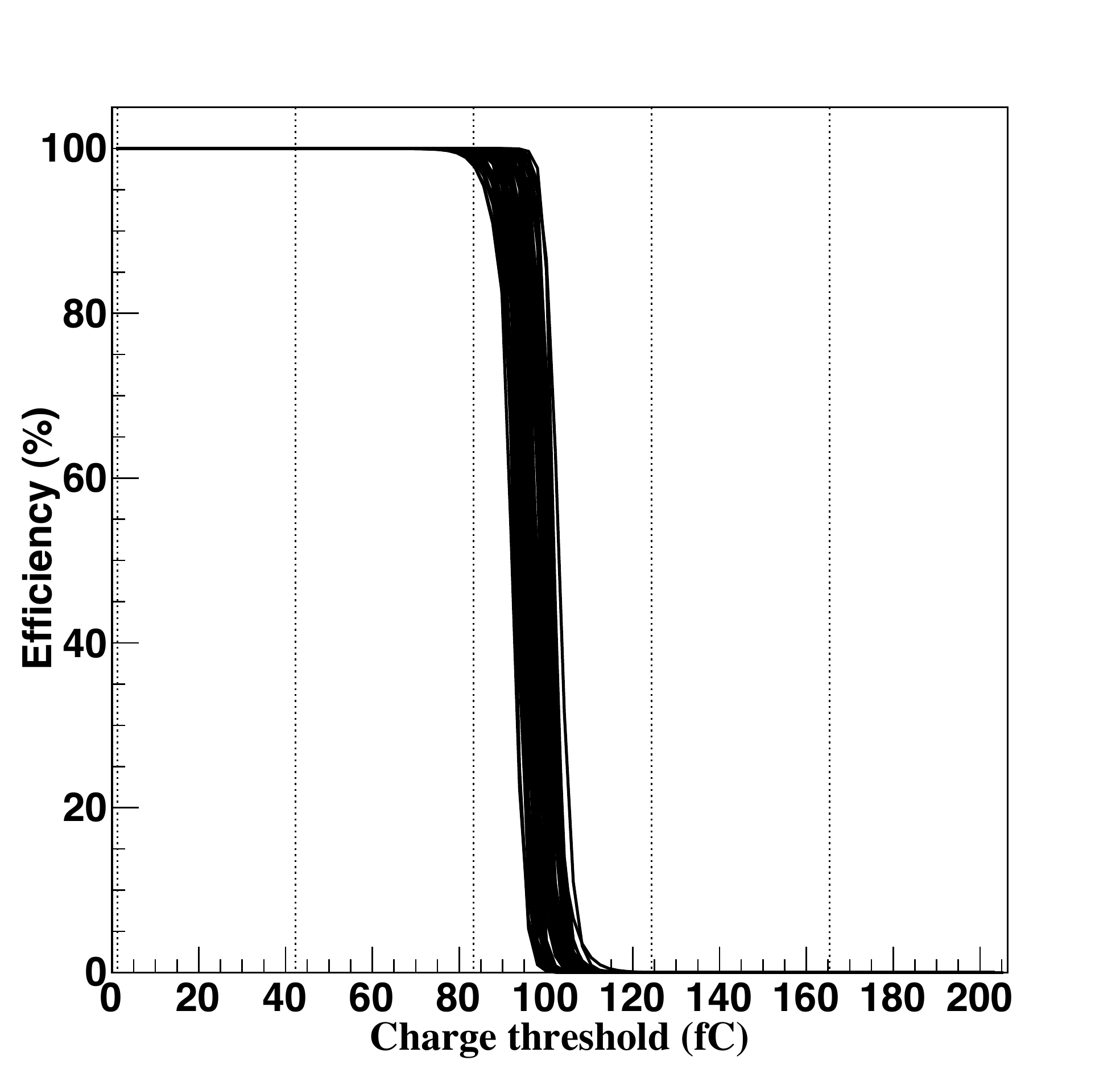}
   \end{centering}
 \end{minipage}
\caption{Efficiency curves before (left) and after correction (right) for 64 channels (one ASIC).}
\label{fig.scurve} 
 \end{center}
\end{figure}

\begin{figure}[h]
 \begin{center}
 \begin{minipage}{0.475\textwidth}
   \begin{centering}
     \includegraphics[width=1\textwidth]{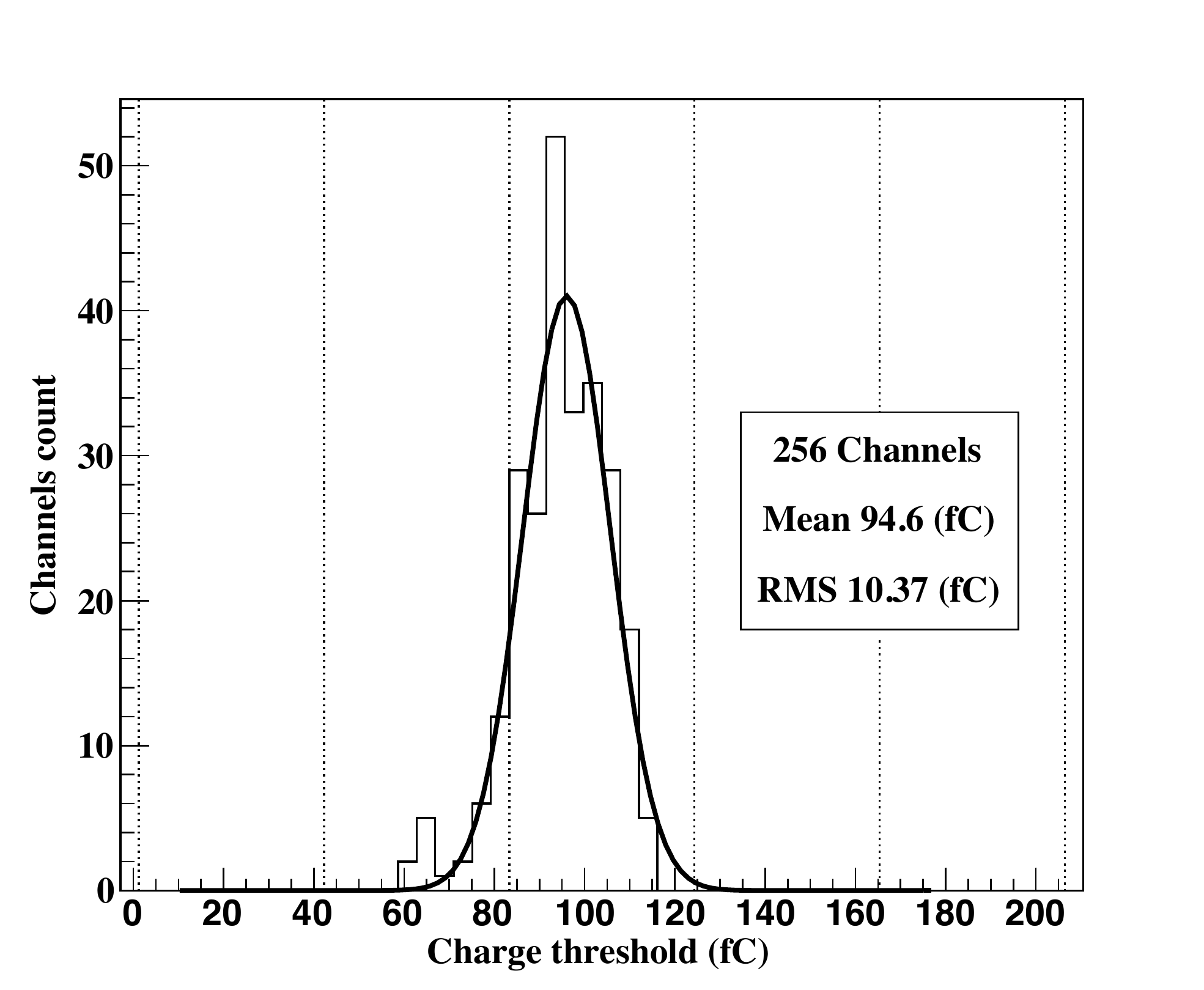}
   \end{centering}
 \end{minipage}
 \begin{minipage}{0.05\textwidth}
 \end{minipage}
 \begin{minipage}{0.475\textwidth}
   \begin{centering}
    \includegraphics[width=1\textwidth]{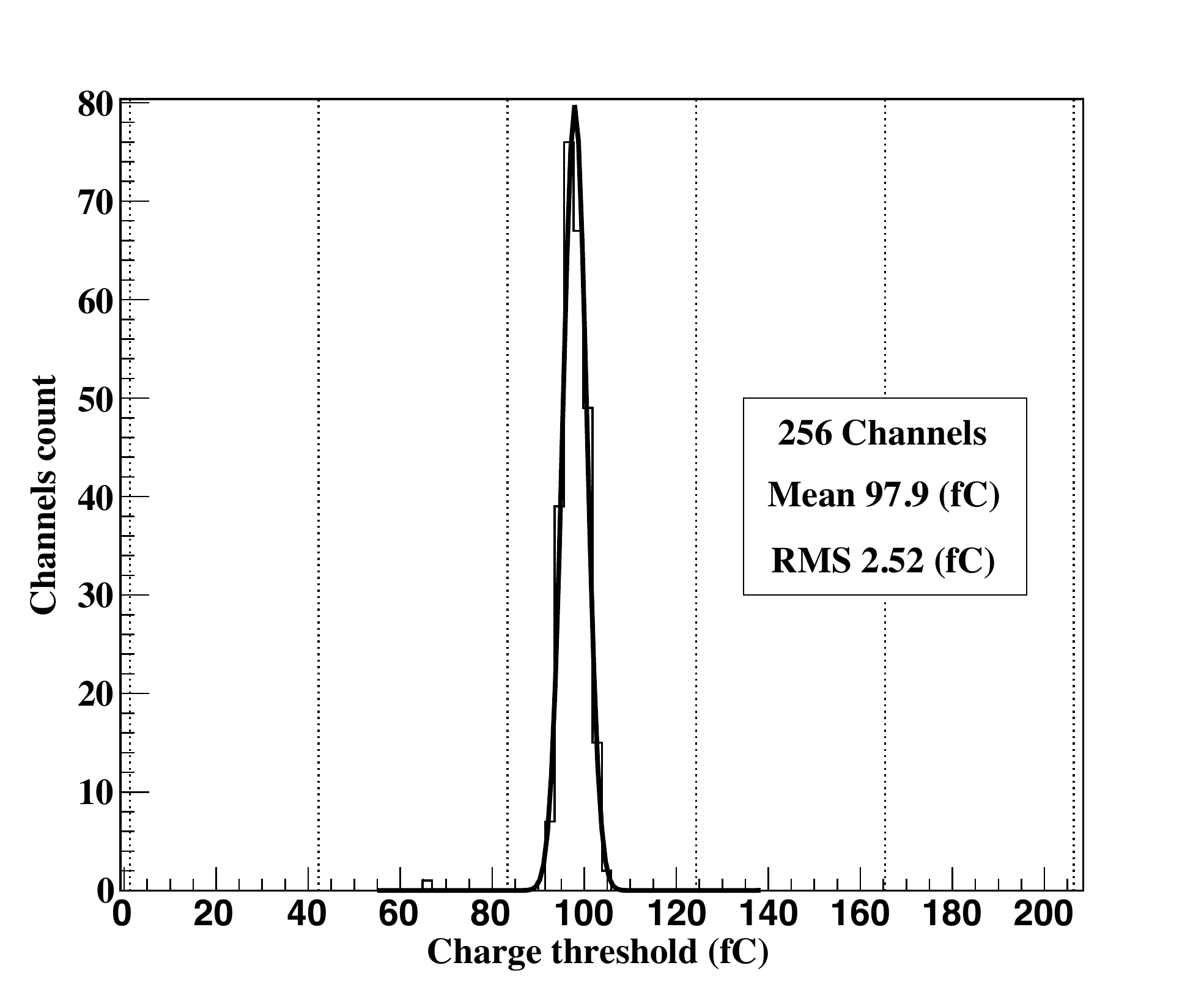}
   \end{centering}
 \end{minipage}
 \caption{Distribution of the 256 inflection points (1 board with 4 ASIC) before and after correction.}
 \label{fig.scurve_distrib} 
 \end{center}
\end{figure}

The gain correction procedure consists in injecting a given charge on each of the 64 channels through the built-in capacitor.  %
The threshold is varied over the whole dynamic range (1024) in steps of 1 DAC unit.  %
The injection is repeated 100 times for each step.  %
This allows the channel response efficiency to be estimated in terms of the applied threshold.  %
The procedure is applied for four different gains.  %
The inflection point on the efficiency curve for each channel and each gain is determined using an error function fit.  %
The inflection point versus gain curves allow individual pad gains to be chosen to minimise the dispersion of the channel response for a given charge. %
The result of this procedure can be seen in Figure~\ref{fig.scurve} and  Figure~\ref{fig.scurve_distrib}, where the gain correction was applied for a 100\u{fC} reference charge.  %
The dispersion of the channels response after gain correction is as low as 2.5\u{fC}. %
The small offset of the mean value is due to the constraint on gain adjustment.  %
The gain value is coded in 6 bits from gain 0 to 4 in 64 steps. 

\newpage

\section{Setup and testbeam description}

\label{sec.testbeam}

Four instrumented GRPCs were exposed to the T10 PS beam at CERN. GRPCs were fixed in a mechanical structure allowing the distance between two successive chambers to be adjusted. %
Scintillation counters placed upstream and downstream of the setup provided the external trigger.  %

\begin{figure}[h]
\begin{center}
\includegraphics[width=0.6\textwidth]{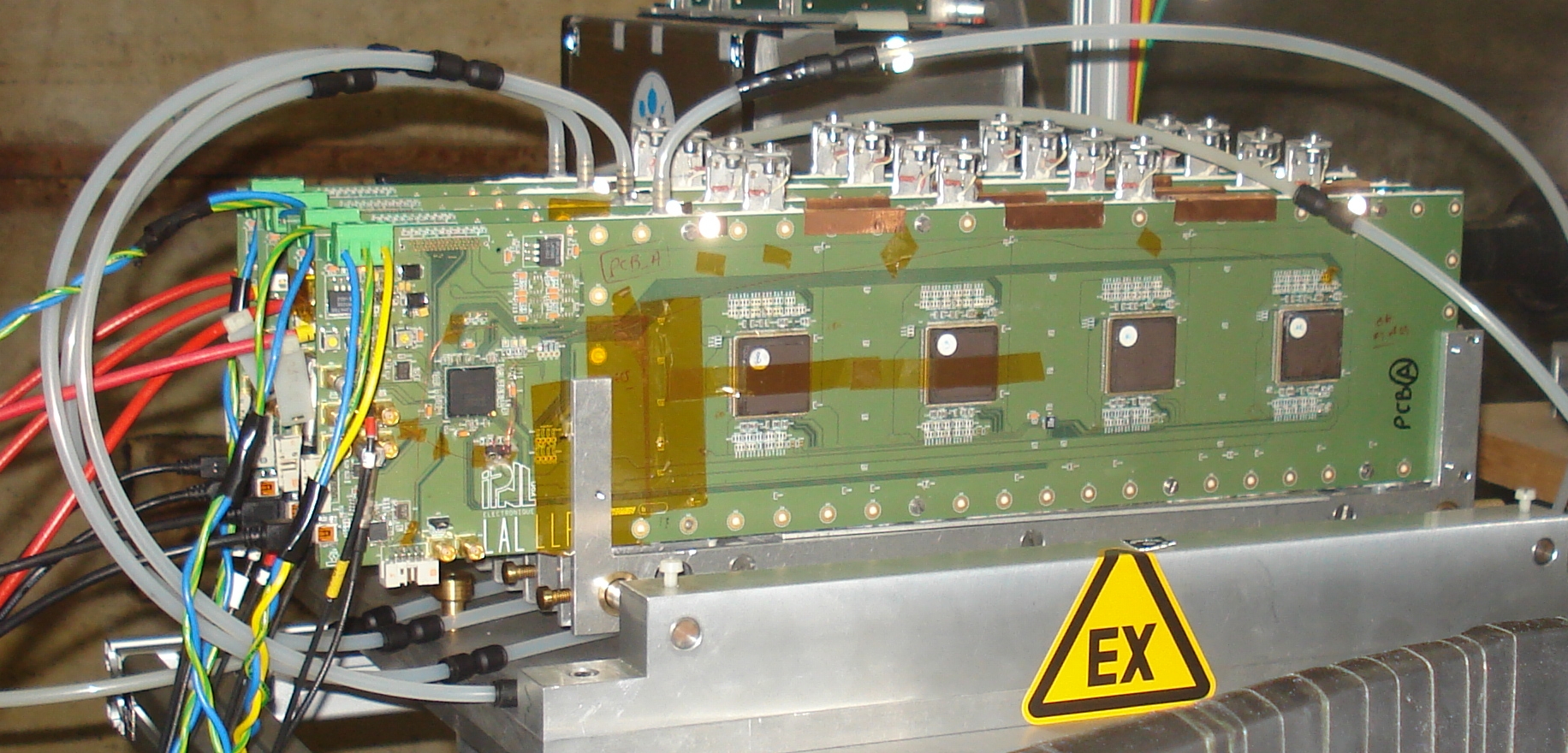}
\caption{Test-beam set-up}
\label{default}
\end{center}
\end{figure}

The beam was essentially composed of pions with a small contamination of muons and electrons.  %
The beam energy could be chosen between 1 and 12\u{GeV}.  
Above 5\u{GeV}, the electron contamination is negligible.  %
Both pions and muons were used in the efficiency study.  
The acquisition system was run in the "triggered mode" (cf section \ref{sec.acquisition}).  %
Hits recorded in the interval between 0 and 400\u{ns} before the external trigger are considered to belong to the particle responsible for the trigger.   %
The length of the time interval takes into account the precision of the 5\u{MHz} internal clock of the ASIC.
Once the trigger-related event is built using the four GRPCs, the earlier frames recorded in memory are sorted according to their time stamp and assembled to build additional events. Those events correspond to earlier particles in the bunch that passed outside the geometrical acceptance of the trigger scintillators. 

\section{Detector performance}

\subsection{Data Quality}
\label{sec.quality}
\begin{figure}[hhh]
\begin{center}
\includegraphics[width=0.5\textwidth]{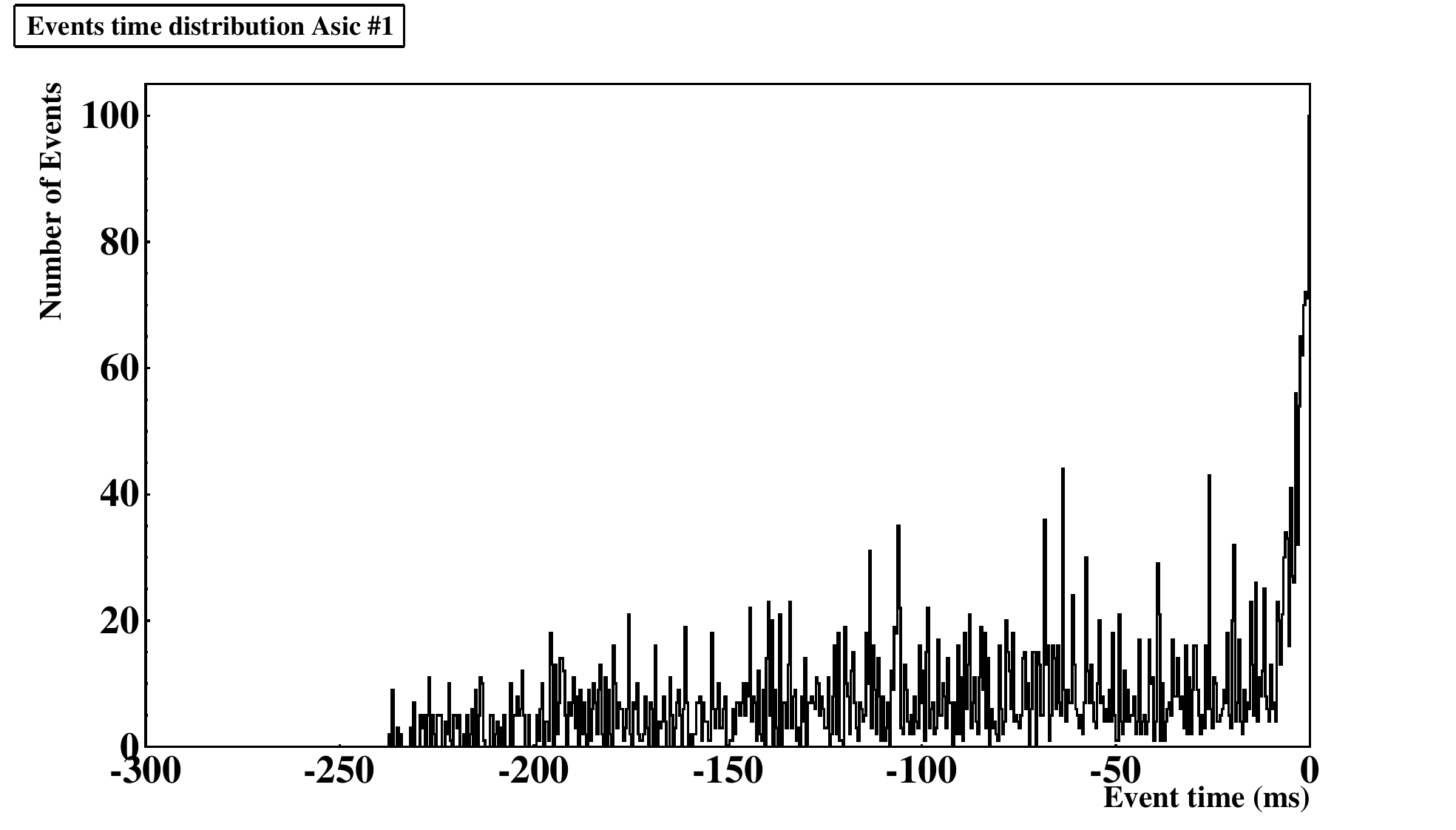}
\caption{Time structure in one ASIC for 100 triggers. The origin (right) represents the trigger time in acquisition.}
\label{fig.sigtime}
\end{center}
\end{figure}

The typical time structure of the signal collected in one ASIC can be seen in Figure~\ref{fig.sigtime}, in which 100 recorded triggers are shown.  %
The origin (right) is the trigger time in acquisition (corresponding to the last frame) while the minimum time represents the first recorded  frame in memory.  %
The peak in the distribution correspond to triggering hits.  The non uniformity of the distribution before the trigger is due to the memory resets of the ASICs that truncate the noise distribution at a different time before each trigger.  

\begin{figure}
\begin{center}
\includegraphics[width=0.7\textwidth]{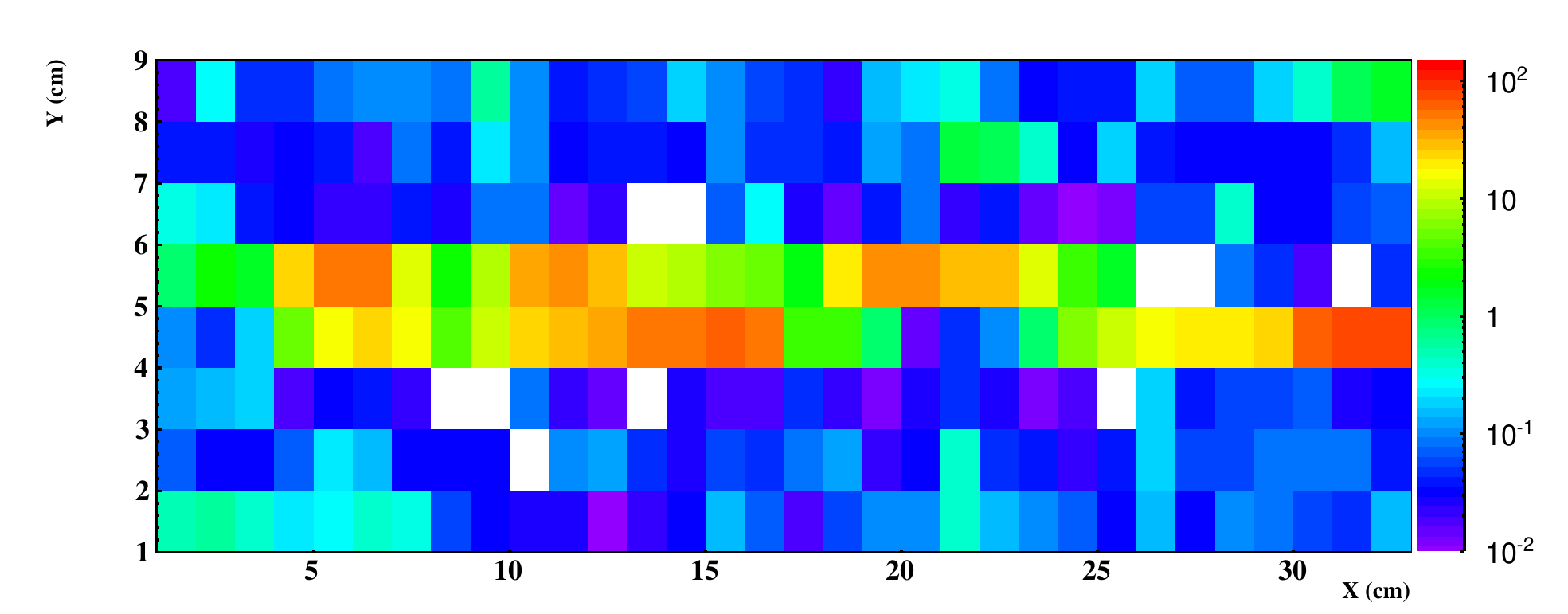}
\caption{ Noise rate map ($\mathrm{Hz/cm^2}$) for one GRPC. The fishing line can be seen in the middle of the GRPC.}
 \label{fig.noiseuniformity}
 \end{center}
\end{figure}

\begin{figure}[h]
 \begin{center}
 \begin{minipage}{0.475\textwidth}
   \begin{centering}
     \includegraphics[width=1\textwidth]{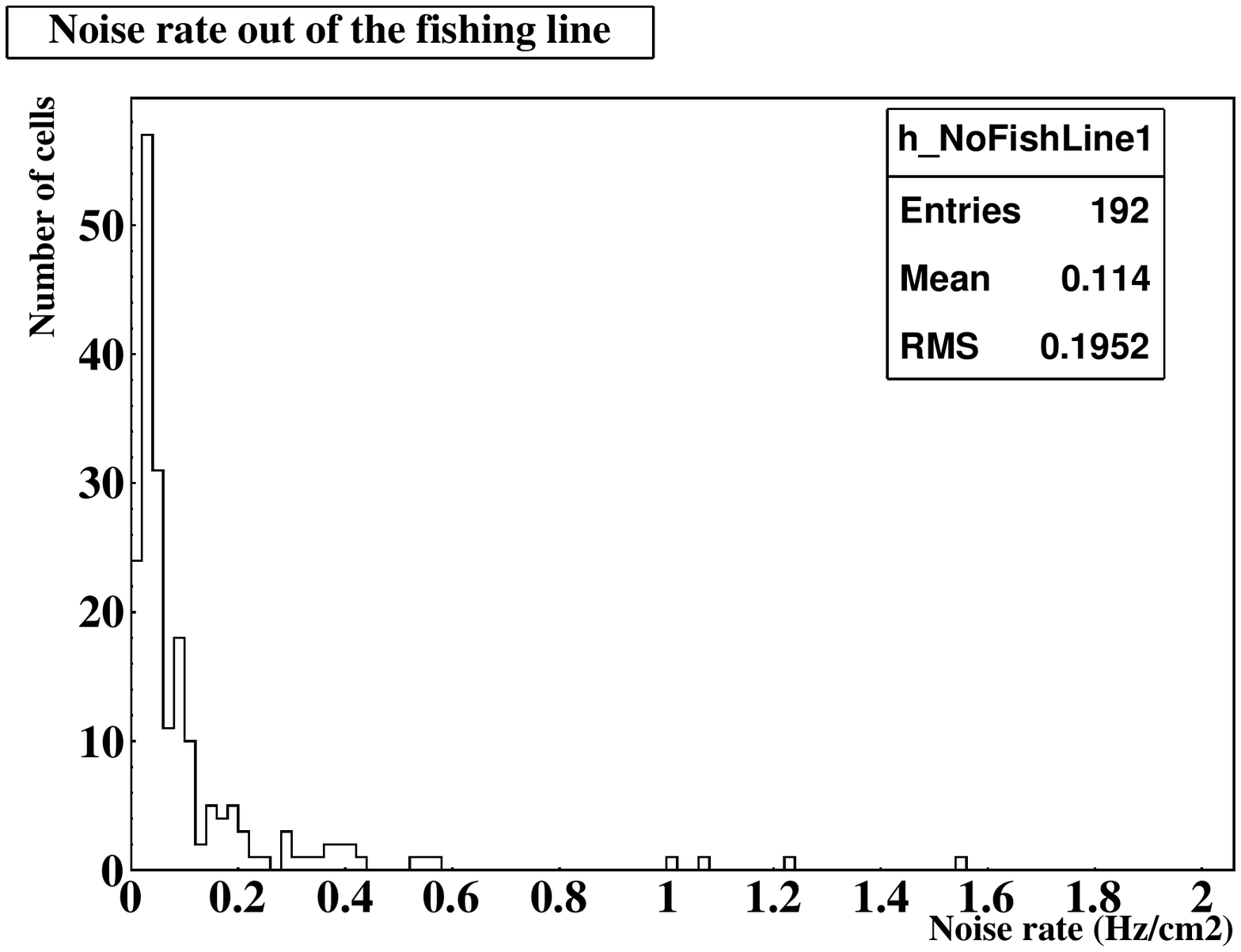}
   \end{centering}
 \end{minipage}
 \begin{minipage}{0.05\textwidth}
 \end{minipage}
 \begin{minipage}{0.475\textwidth}
   \begin{centering}
    \includegraphics[width=1\textwidth]{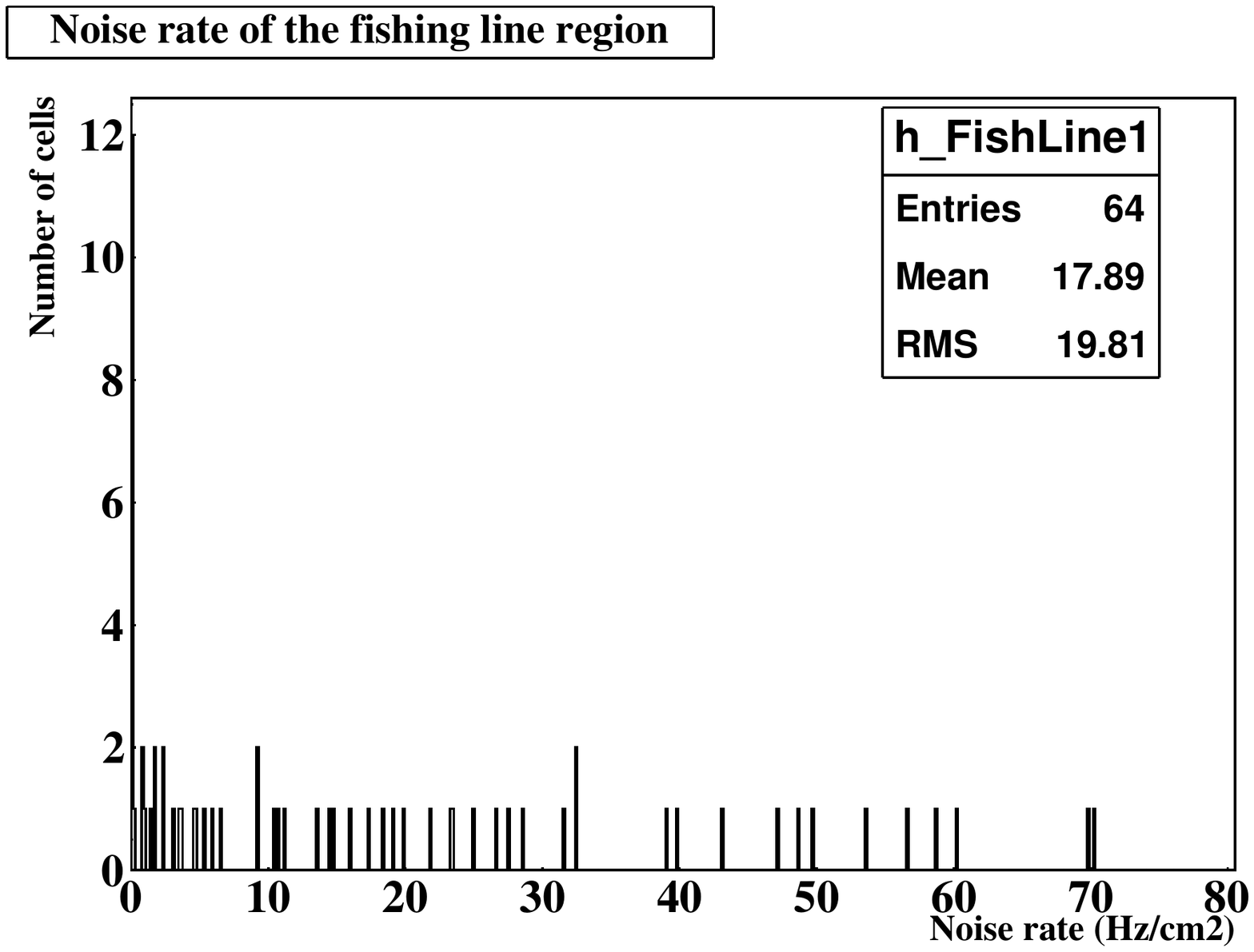}
   \end{centering}
 \end{minipage}
\caption{Noise rate distribution out of the fishing line region (left) and in the fishing line region (right) for 256 channels (4 ASICs).}
\label{fig.noiserate} 
 \end{center}
\end{figure}

Using a random external trigger, the noise rate was measured in a GRPC with a threshold of 165\u{fC} and a polarisation voltage of 7.4\u{kV}. %
Its spatial noise rate distribution is shown in Figure~\ref{fig.noiseuniformity}.  %
Out of the fishing line region, the average noise rate is measured to be 0.11\u{Hz/cm^2}. %
The noise rate distribution in and out of the fishing line zone can be seen in Figure~\ref{fig.noiserate}.  %
With a 400\u{ns} time window used for event reconstruction, the fake hit rate is about $4.4\times10^{-8}\u{cm^{-2}/event}$. %

The fraction of dead channels was measured to be $<0.1\u{\%}$ (1 out of 1024). %
Noisy channels with high pedestal values (with respect to the others) represented only $1\u{\%}$ of the total number of channels. They could be neutralised applying an overall threshold rise to the corresponding ASICs.  %


\subsection{Tracking}
\label{sec.tracking}
The efficiency and the multiplicity of each of the 4~GRPCs in the setup were studied using  tracks built with the hits in the other GRPCs. Every pair of  hits belonging to two different GRPCs (except the studied one) was used to build a track candidate.  The track is kept if at least one hit is found in the third GRPC within a 2\u{cm}-radius around the expected impact position. When different track candidates are present, the one with the best $\chi^2$ is selected.
Hits are then searched for in the studied GRPC around the expected impact position of the selected track.  %
The efficiency is determined by the presence of at least one hit within a 2\u{cm}-radius area around the projected impact point.  %
The multiplicity is defined as the number of fired pads within a 2\u{cm}-radius around the found pad.

\subsection{Standard graphite-coated GRPC}

\subsubsection{Threshold dependence}
The efficiency dependence upon the digital threshold was studied at the 7.4\u{kV} working point (see Figure \ref{fig.EffAngle} (left)) for thresholds above the pedestal. 
Based on this study, the trigger threshold was optimised to maximise the efficiency while minimising the  noise contribution.  %
A value of 165\u{fC} was chosen, corresponding to about one tenth of the mean MIP charge (measured using a standard analog read-out). %

\subsubsection{Angle dependence}
\begin{figure}[h!]
\begin{center}
\begin{tabular}{cc}
\includegraphics[width=0.5\textwidth]{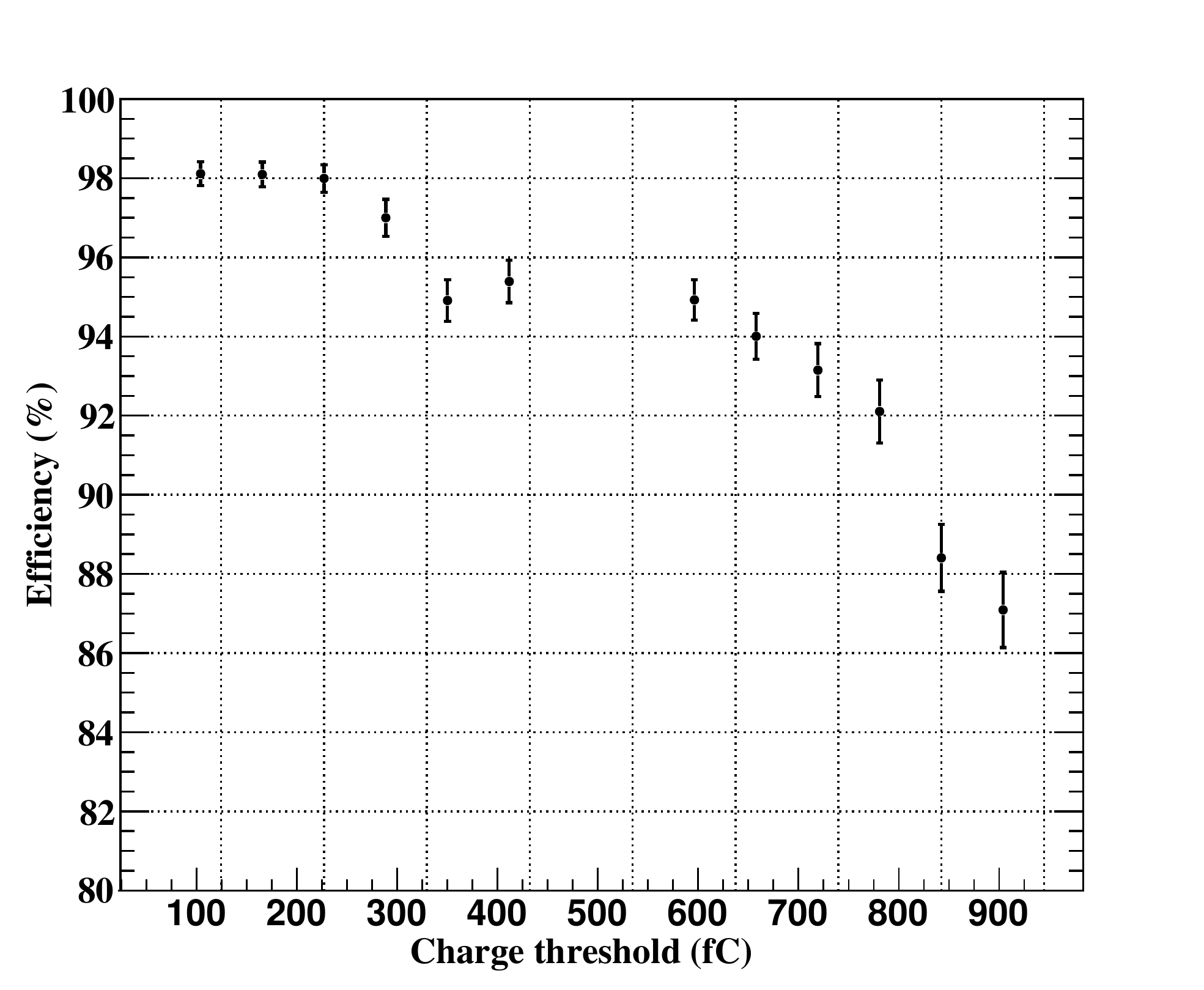}
&\includegraphics[width=0.5\textwidth]{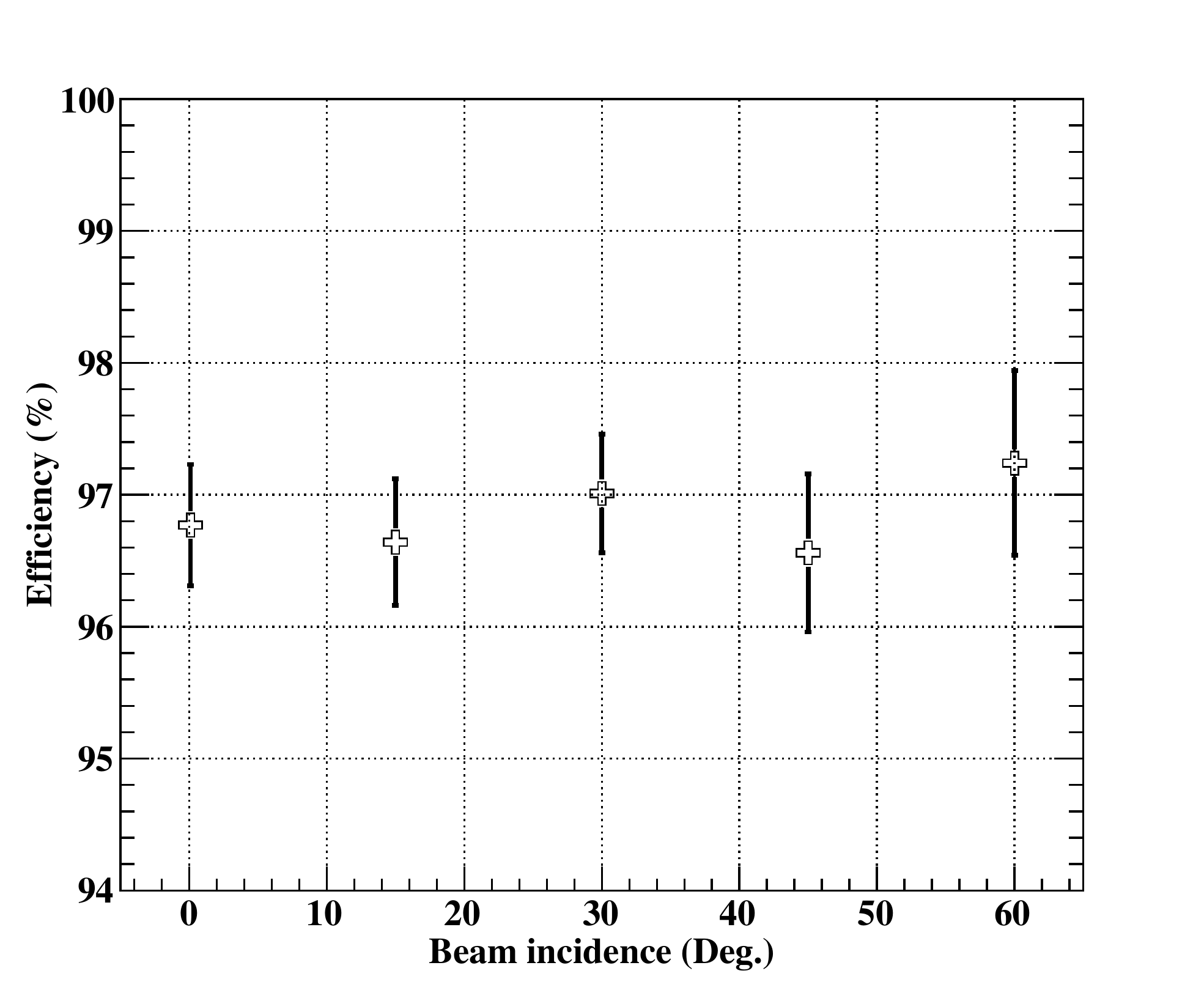}
\end{tabular}
\caption{Efficiency as a function of threshold (left) and  angle (right), graphite-coated chambers \label{fig.EffAngle}}
\end{center}
\end{figure}

Charge multiplication in our detector occurs in a 1.2\u{mm} thin gas gap. The efficiency was observed to be independent of the incidence angle between the incoming particle  and the normal to the detectors up to $60^{\circ}$  as can be seen in Figure \ref{fig.EffAngle} (right). This observation enhances the geometrical flexibility for future ILC calorimeter design.

\newpage
\subsubsection{ GRPC performance with high voltage}
\label{sec.HV1}
\begin{figure}[h]
\begin{center}
\begin{tabular}{cc}
\includegraphics[width=0.5\textwidth]{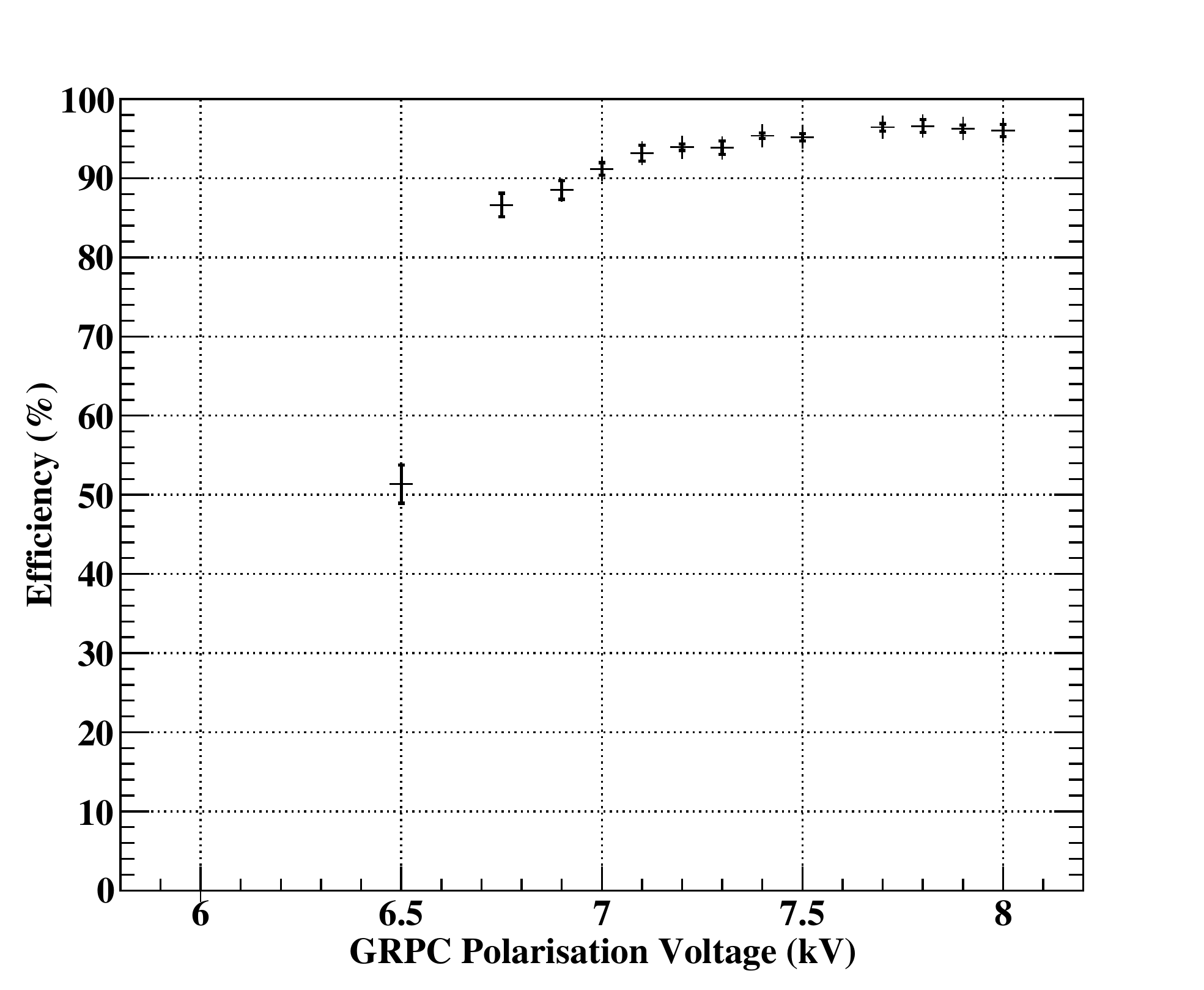}
&
\includegraphics[width=0.5\textwidth]{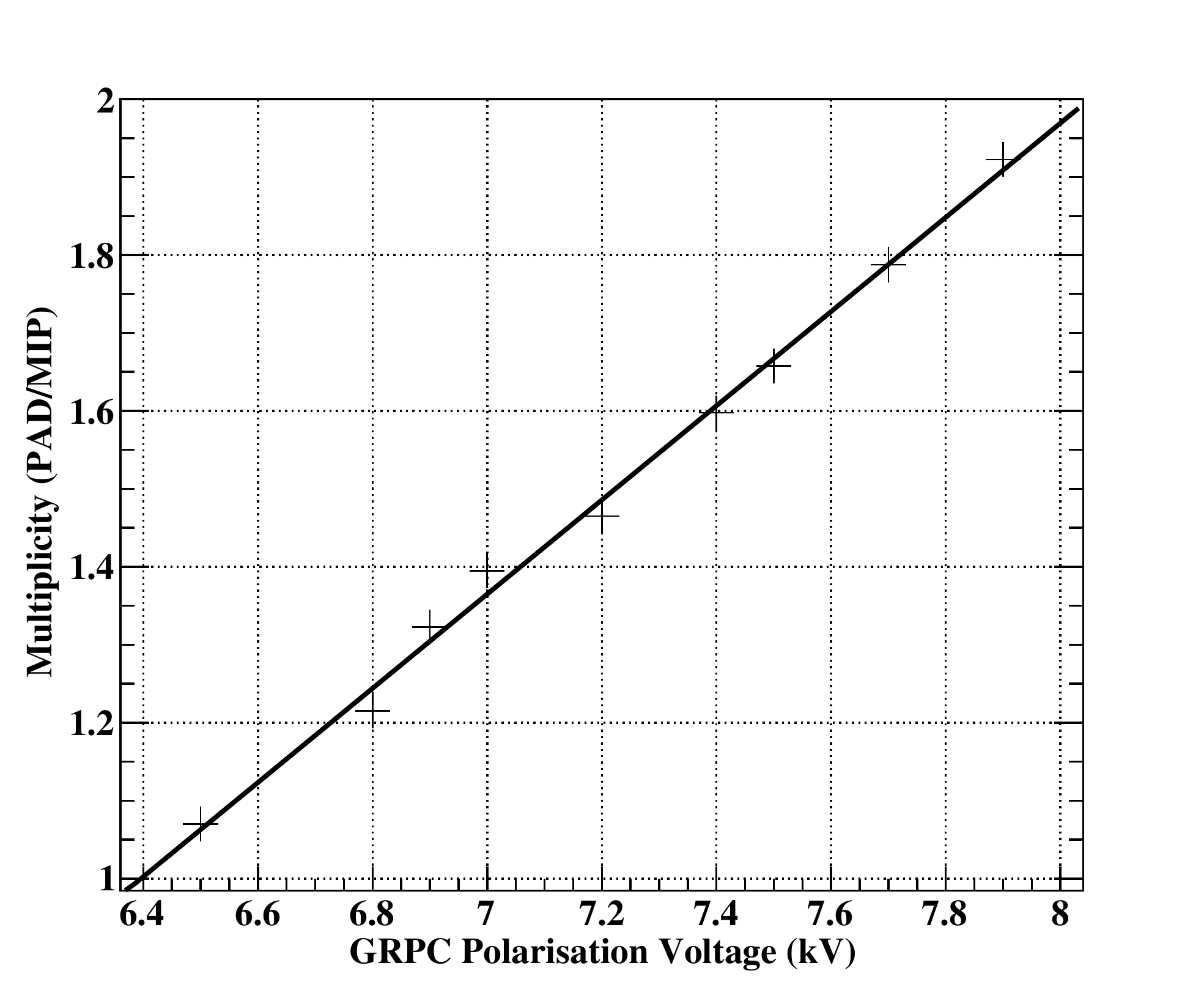}
\end{tabular}
\caption{Average efficiency (left) and multiplicity (right) as a function of high voltage  for standard graphite-coated GRPCs (threshold:165 fC).}
\label{fig.MultEffHVmoy}
\end{center}
\end{figure}

A detailed study of the behaviour of the detectors with high voltage was performed using MIPs. According to Figure \ref{fig.MultEffHVmoy} (left) the charge multiplication process starts to become detectable in our detectors at about 6.5~kV. The efficiency then rises and reaches a plateau at 7.3 kV. Figure \ref{fig.MultEffHVmoy} (right) shows that the number of fired pads increases steadily with the high voltage in the studied range. The increase can be explained by the fact that the larger total charge produced in the avalanche induces more signal  above the threshold in the neighbouring pads.

Large multiplicity values could affect track separation and reduce the performance of the future PFA-oriented hadronic calorimeter.  %
A compromise is to be found between high efficiency and low pad multiplicity.  %
A polarisation of 7.4\u{kV} was chosen as the working point for our detectors.  %
The choice of 7.4 rather than 7.3 is to avoid significant efficiency variations in case of small electric field variations which may be caused by a slight local change of the distance between the two resistive plates. Deformation of glass plates were calculated to be less than 50\u{\umu m}. That leads to a maximum  variation of 4\% of the electric field used for charge multiplication.

\begin{figure}[h!]
\begin{center}
\begin{tabular}{cc}
\includegraphics[width=0.5\textwidth]{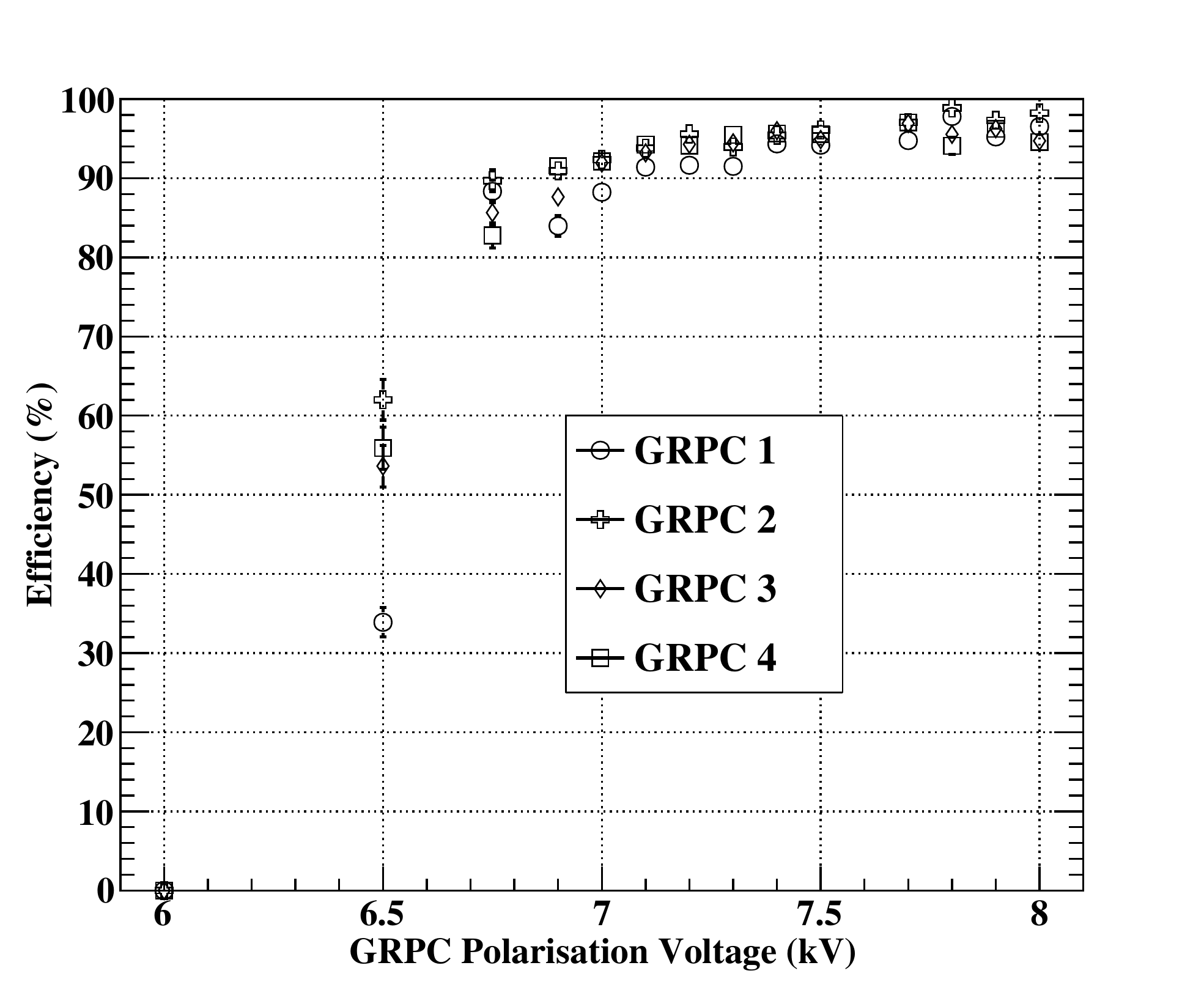}
&\includegraphics[width=0.5\textwidth]{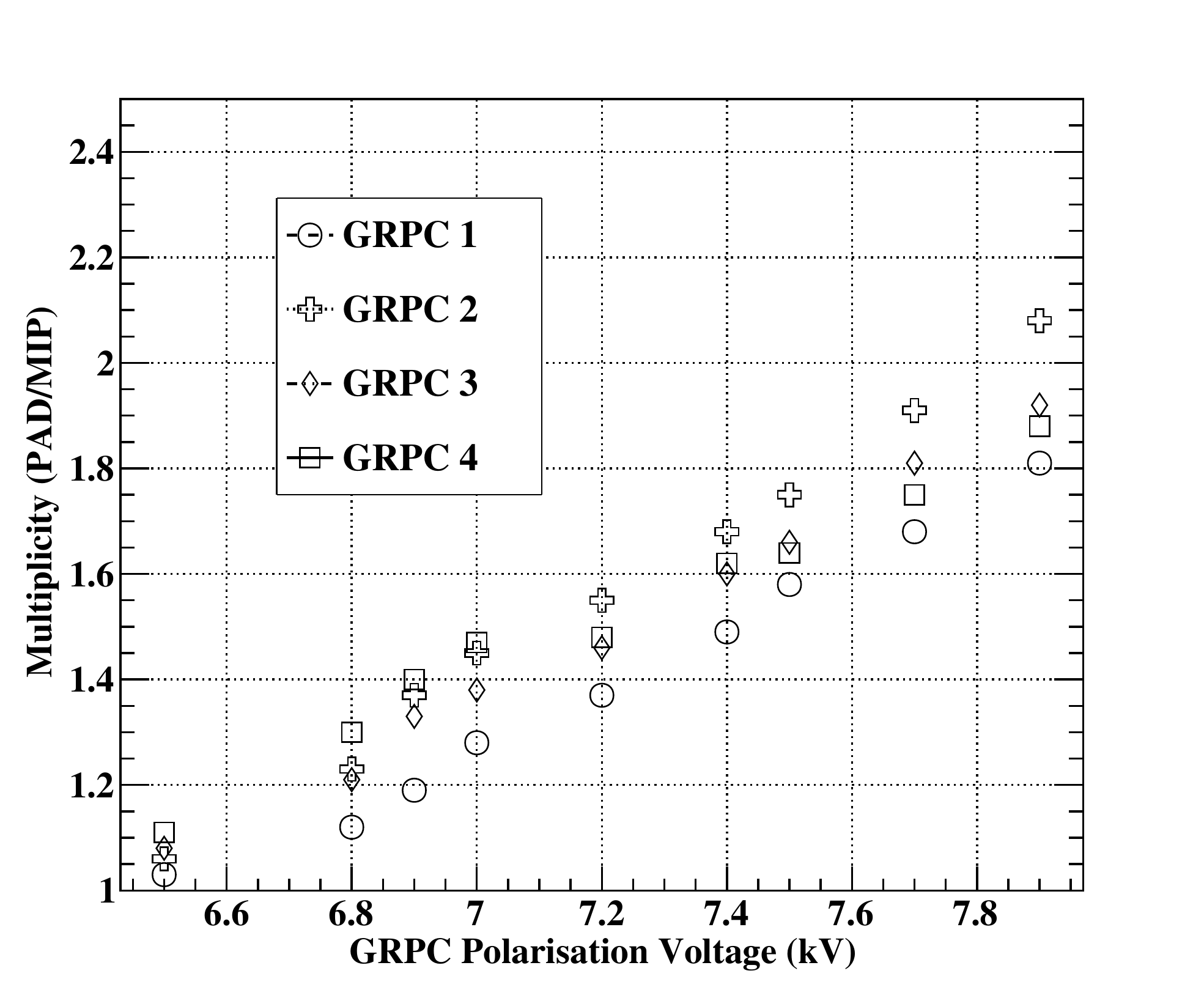}
\end{tabular}
\caption{Efficiency (left) and multiplicity (right) as  function of  high voltage for the four graphite-coated chambers}
\label{fig.MultEffHV4graphite}. 
\end{center}
\end{figure}

The uniformity of the high voltage response of the four standard graphite-coated  GRPCs built using the same technique is shown in Figure \ref{fig.MultEffHV4graphite}.  
This reproducibility is a strong argument in favour of using GRPCs as the hadronic calorimeter's sensitive medium.

\newpage
\subsubsection{Chamber homogeneity}
An important feature of gas detectors is their homogeneity.  %
The PS beam optics were
adjusted to cover the whole detector surface to assess the homogeneity of our detector.  %

High statistics needed for efficiency and multiplicity maps were obtained by performing a full
reconstruction of the ASIC's history prior to the trigger using the internal memory of up to 128 frames. Hits  are
clustered in events if their time difference is below 400\u{ns}.
Efficiency and multiplicity maps of each detector were obtained at the bias voltage of 7.4\u{kV} using the same method described in section~\ref{sec.tracking}.

\begin{figure}[h!]
\begin{center}

  \includegraphics[width=0.8\textwidth]{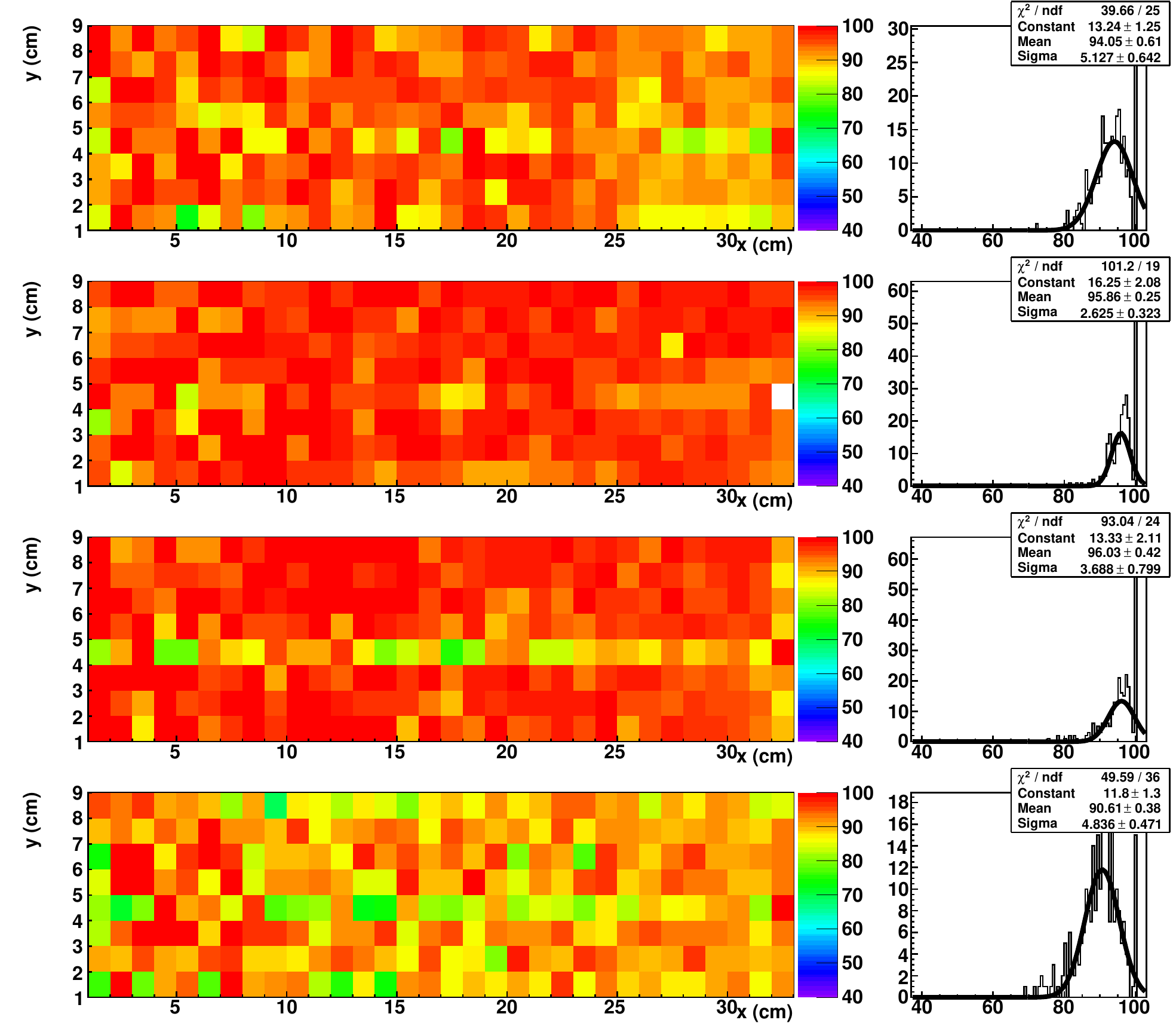}
  \caption{Efficiency maps (left ) of 4 GRPCs for a 7.4\u{kV} bias voltage
    and distributions (right) of the values for the 256 pads of each GRPC,
    fitted by a gaussian distribution.}
  \label{fig.Effmap}
\end{center}
\end{figure}
\begin{figure}[h]
 \begin{center}
 
     \includegraphics[width=1\textwidth,height=12 cm]{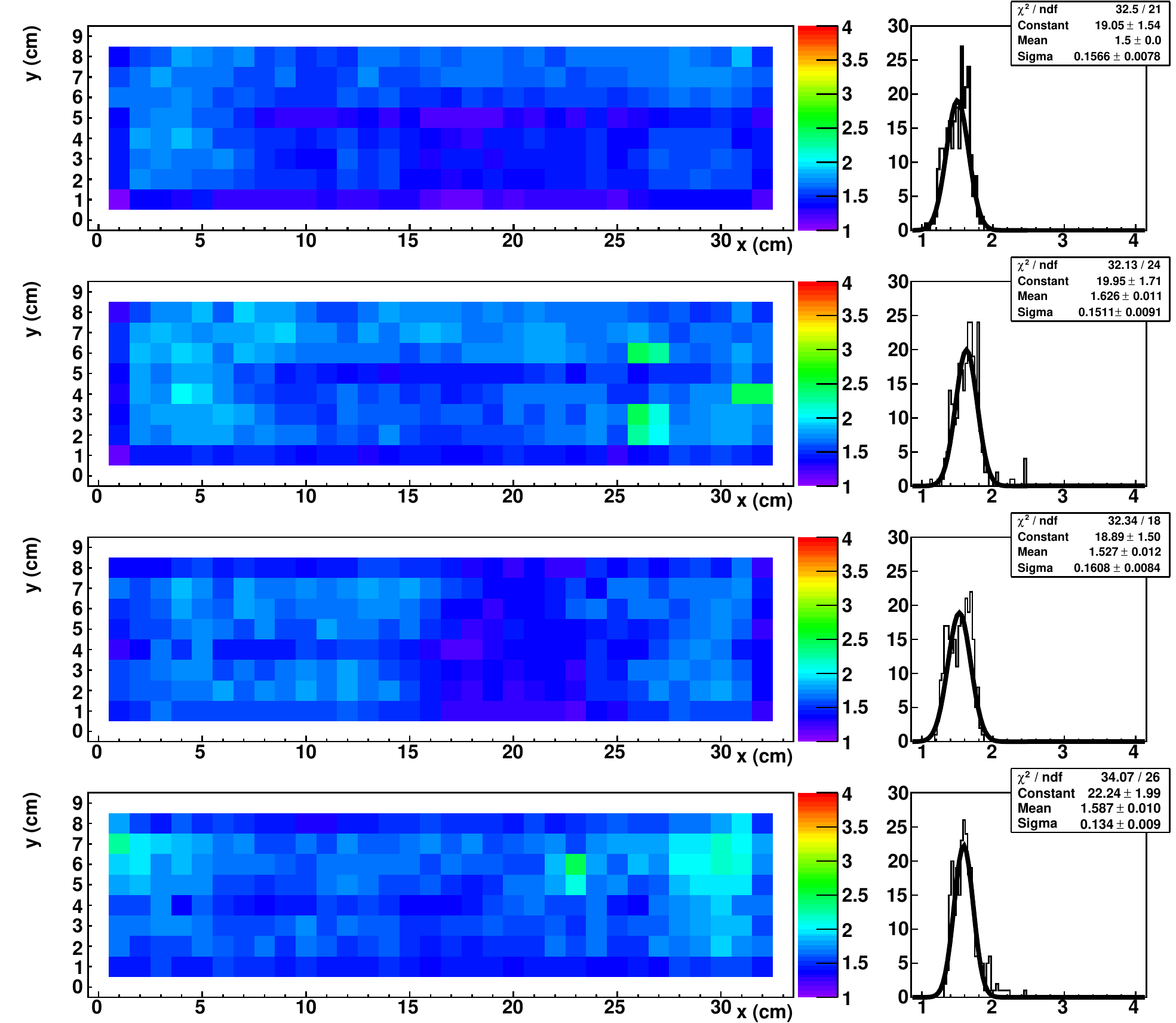}
   \caption{Multiplicity maps (left) of 4 GRPCs for a 7.4\u{kV} bias
    voltage and distributions (right) of the values for the 256 pads of each
    GRPC, fitted by a  gaussian distribution.}
  \label{fig.MultMap}
 \end{center}
\end{figure}

The main source of variations of the efficiency and multiplicity are  the fishing line (crossing the chambers longitudinally in the middle) and the
borders as can be observed on Figures~\ref{fig.Effmap} and \ref{fig.MultMap}.
No structure can be seen  in the efficiency map away from the edges and fishing lines.  A global
spread, including the statistical error ($\approx$ 100 events/pad), of $3\u{\%}$ was
observed inside single chambers and between different chambers. 
The slightly reduced efficiency of the fourth chamber with respect the other three is due to a  mechanical fixation problem  during the homogeneity test, that created a small gap between the electronics board and the detector. 
The multiplicity maps present a globally uniform pattern with a 
Gaussianly-distributed spread of $\sim0.15$.  One region (3/4 right of
the third chamber from the top) presents a slightly lower multiplicity and efficiency. This is due to a higher threshold setting of the corresponding ASIC.
Cross talk was observed between 3 pairs of pads in the second chamber from the top. This shows up as 3 higher multiplicity regions of the chamber.

\newpage
\subsubsection{Interpad effects}
The avalanche spread at the anode level, originating from a charged particle crossing a 2~mm-gas-gap GRPC operated in the avalanche mode is estimated to be 1-2~mm \cite{Chinese}.
This extension is expected to be reduced in our 1.2~mm~gas-gap GRPC.  %
Therefore, the distance separating two pads was chosen to be 500\u{\umu m} to keep our detector sensitive to particles crossing between two adjacent pads.

The effect of this gap was checked using the track prediction of one of the two arms of the EUDET pixel telescope~\cite{eudet}.  %
The telescope was placed at a distance of 20\u{cm} in front of our setup during the testbeam at the PS.  %
Due to the limited sensitive area of the EUDET telescope ($7\times7\u{mm^2}$), our setup was placed on a movable table in order to cover a larger area of our detector. %
Beam tracks were recorded for different positions of our setup with respect to the telescope position. 

The first horizontal position was chosen in such a way that the tracks seen in the telescope were on both sides of  the right vertical edge of our detector.  The vertical position corresponded to the center of a pad row. 
Relative positions of both setups were measured precisely using photogrammetry.  %
This method yields a distance resolution of few microns.  %
In addition, the photogrammetry spots were chosen on the two setups so that the uncertainty of the EUDET telescope track relative precision with respect to our GRPC detector is limited to the mechanical precisions of both the EUDET telescope and our electronics board.  %
The combined uncertainty is estimated to be lower than 100 microns which is five times smaller than the interpad distance.  %
Our setup was then moved in steps of 3.5\u{mm} using a micrometrical precision translator.  %
In our detector, this method guarantees a 50\u{\%} overlap of the studied zones in two successive steps.  %
The number of steps was optimised to cover at least two interpad regions. 
The precision of the prediction of particle impact points in the GRPCs was smaller than 10\u{\umu m}.

Although the two setups used different acquisition systems, a unique external trigger was used to record events in both. External triggers were accepted if none of the two systems were busy. This allows to unambiguously associate events in the two systems.  

The tracks built using the EUDET telescope were projected onto the closest GRPC of our setup and their impact points were predicted. Tracks in the GRPC setup is built if at least one hit is found in the edge region of two different GRPC detectors excluding the studied one. The tracks matched in both setups were used.

\begin{figure}[h]
\begin{center}
\includegraphics[width=0.6\textwidth]{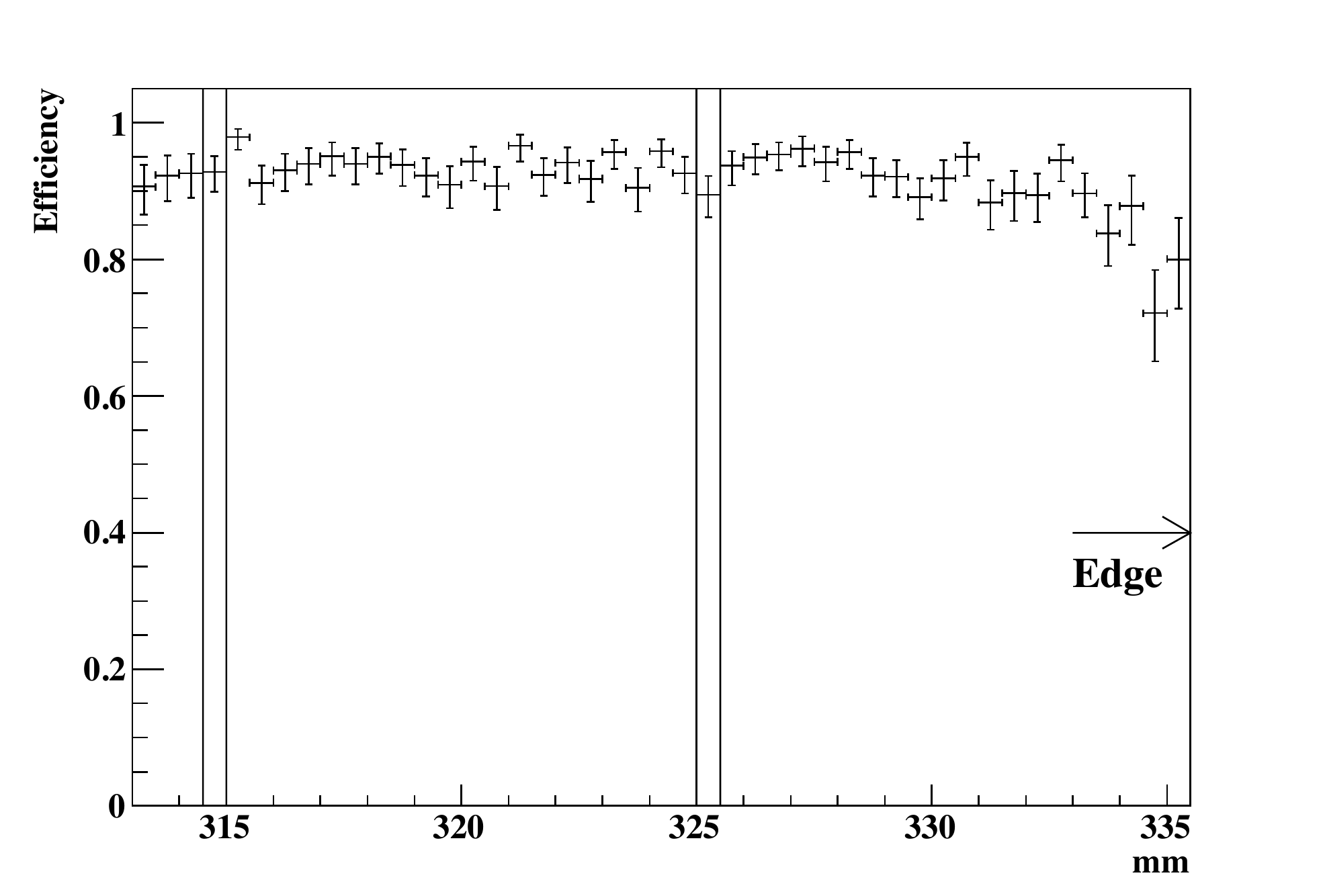}
\caption{Efficiency along the copper pads (the vertical lines indicate interpad positions).}
\label{fig.interpad}
\end{center}
\end{figure}

Hit pads surrounding the projected impact point are then looked for and the track detection efficiency is estimated. 
The efficiency as a function of the position along the horizontal side of a GRPC is shown in Figure~\ref{fig.interpad}.  %
The two interpad positions as expected from the photogrammetry are shown on the same figure.  %
No significant change in efficiency is observed in the two interpad zones.  %
The statement is valid for all the zones 3\u{mm} away from the detector edge.  %

The same study shows a 15\u{\%} lower efficiency at the edge up to 3\u{mm} inside the detector.  %
This reduction is probably due to the electric field deformation near the chamber edge.  %
It is however necessary to investigate this effect in more detail in order to improve the detector performance.

\subsection{High-resistivity coatings} 
\begin{figure}[h]
\begin{center}
\begin{tabular} {cc}
\includegraphics[width=0.49\textwidth]{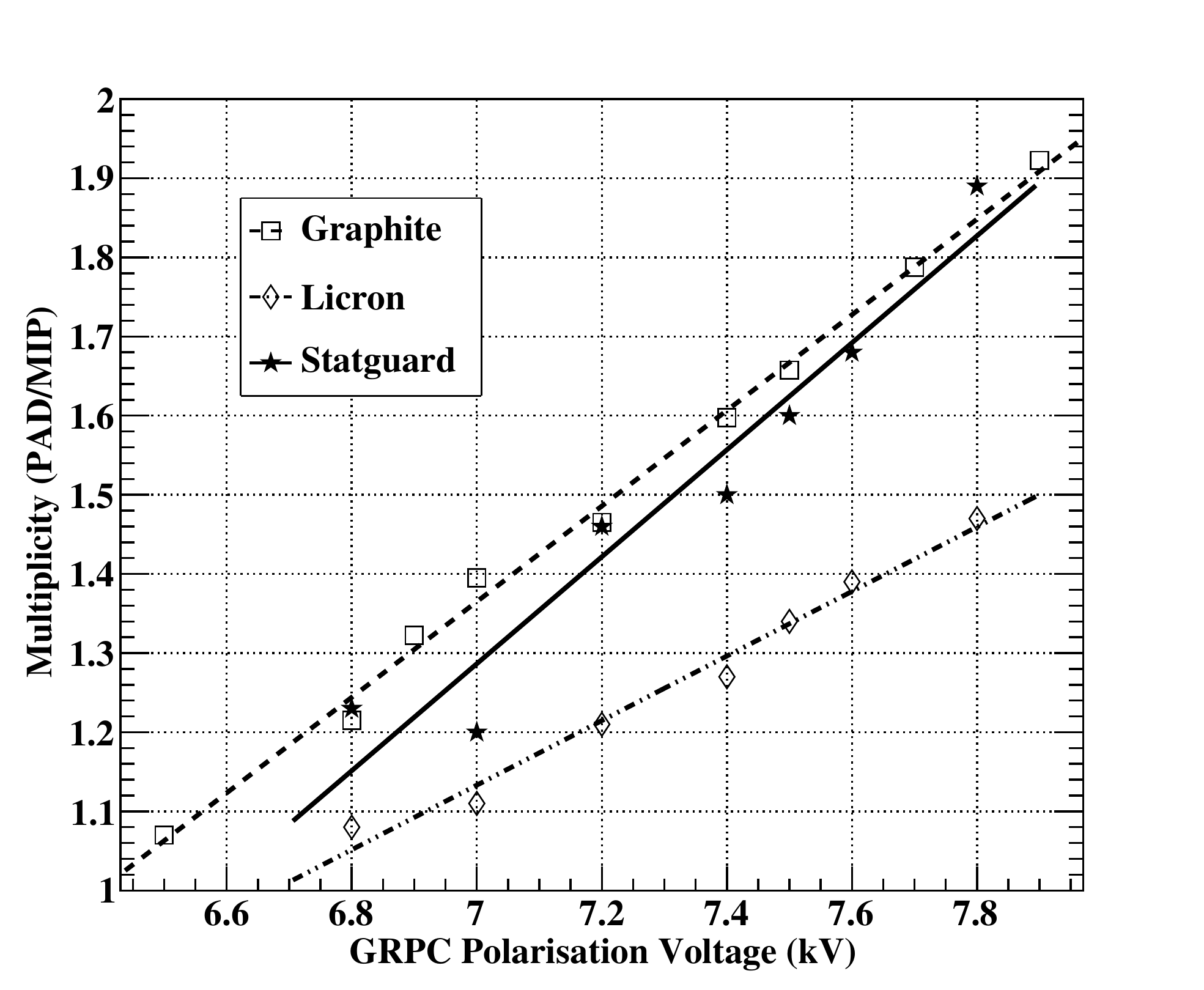}
&
\includegraphics[width=0.49\textwidth]{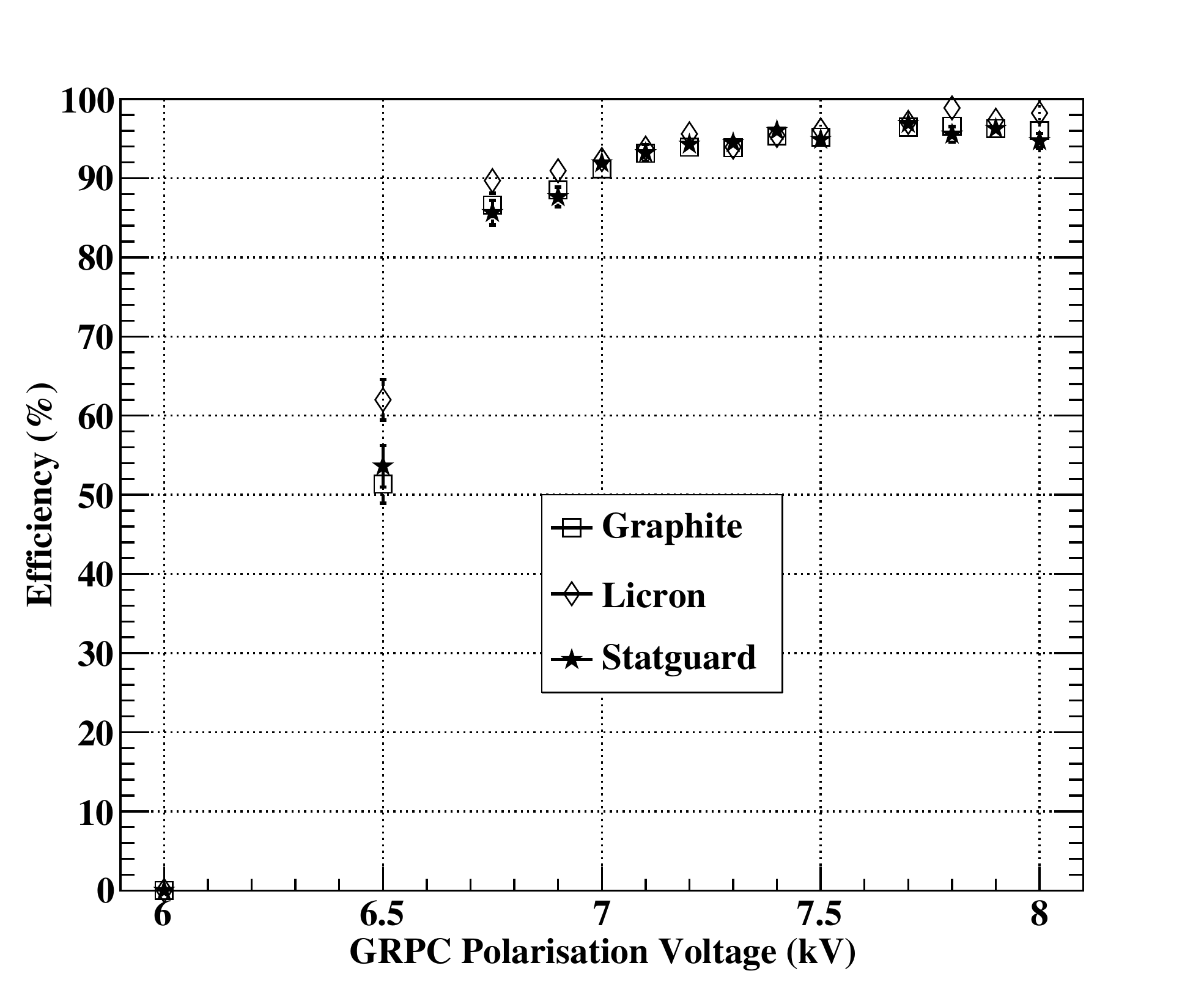}
\end{tabular}
\caption{Multiplicity (left)  and Efficiency (right) as a function of high voltage for graphite, Licron and Statguard coatings.}
\label{fig.MultEffHVcoatings}
\end{center}
\end{figure}

Other coating materials with higher resistivity were tested to improve the detector spatial resolution. The resistivity of the coating (see Table \ref{tab.coating}) has a strong impact on charge dispersion and renewal. The magnitude of this effect is shown in the multiplicity graph of Figure \ref{fig.MultEffHVcoatings} : the higher the resistivity the more localised the observed signal is.
The Licron coating shows the best results in terms of multiplicity.  The validity of the Licron coating option is supported by the fact that all tested coatings show the same efficiency behaviour as a function of the bias voltage.

\section{Conclusion}
We have developed a new approach to the design of thin GRPCs with embedded multi-threshold read-out electronics. Several prototypes have been successfully built and tested in pion beams. All chambers show the required high efficiency, homogeneity,  stability and reproducibility needed for a high-granularity hadronic calorimeter for the ILC.
Low-pad multiplicity was obtained using highly resistive coatings with no incidence on the efficiency.
Although the importance of the 2-bit read-out of our electronics was not emphasized in this work, it is an important feature of our design.  This is supported by ongoing studies based on a realistic simulation that seem to show the importance of having semi-digital read-out electronics to improve the energy resolution at high energy.

\end{document}